\documentclass{article}

\usepackage{arxiv}

\usepackage[utf8]{inputenc} 
\usepackage[T1]{fontenc}    
\usepackage[hidelinks]{hyperref}       
\usepackage{url}            
\usepackage{booktabs}       
\usepackage{amsfonts}       
\usepackage{nicefrac}       
\usepackage{microtype}      
\usepackage{lipsum}		
\usepackage{graphicx}
\usepackage{natbib}
\usepackage{doi}
\usepackage{amsmath}
\usepackage{bm}

\DeclareMathOperator{\DY}{\pi_{\text{DY}}}
\DeclareMathOperator{\CI}{\pi_{\text{CI}}}
\DeclareMathOperator{\HPP}{\pi_{\text{HPP}}}
\DeclareMathOperator{\Cov}{Cov}

\usepackage{xr}
\makeatletter
\newcommand*{\addFileDependency}[1]{
  \typeout{(#1)}
  \@addtofilelist{#1}
  \IfFileExists{#1}{}{\typeout{No file #1.}}
}
\makeatother

\newcommand*{\myexternaldocument}[1]{%
    \externaldocument{#1}%
    \addFileDependency{#1.tex}%
    \addFileDependency{#1.aux}%
}
\myexternaldocument{file2}

\title{A hierarchical prior for generalized linear models based on predictions for the mean response}

\author{ \href{https://orcid.org/0000-0002-6112-9030}{\includegraphics[scale=0.06]{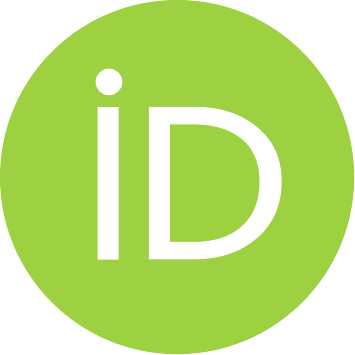}\hspace{1mm}Ethan M. Alt} \\
	Department of Biostatistics\\
	University of North Carolina\\
	Chapel Hill, NC 27516 \\
	\texttt{ethanalt@live.unc.edu} \\
	\And
	Matthew A. Psioda \\
	Department of Biostatistics\\
	University of North Carolina\\
	Chapel Hill, NC 27516 \\
	\texttt{matt\_psioda@unc.edu} \\
	\AND
	Joseph G. Ibrahim \\
	Department of Biostatistics\\
	University of North Carolina\\
	Chapel Hill, NC 27516 \\
	\texttt{ibrahim@bios.unc.edu} \\
}

\renewcommand{\shorttitle}{\textit{arXiv} Template}

\hypersetup{
pdftitle={A template for the arxiv style},
pdfsubject={q-bio.NC, q-bio.QM},
pdfauthor={David S.~Hippocampus, Elias D.~Striatum},
pdfkeywords={First keyword, Second keyword, More},
}

\begin{document}
\maketitle

\begin{abstract}
There has been increased interest in using prior information in statistical analyses.  For example, in rare diseases, it can be difficult to establish treatment efficacy based solely on data from a prospective study due to low sample sizes. To overcome this issue, an informative prior for the treatment effect may be elicited. We develop a novel extension of the conjugate prior of \citet{chen_conjugate_2003} that enables practitioners to elicit a prior prediction for the mean response for generalized linear models, treating the prediction as random. We refer to the hierarchical prior as the hierarchical prediction prior. For i.i.d. settings and the normal linear model, we derive cases for which the hyperprior is a conjugate prior. We also develop an extension of the HPP in situations where summary statistics from a previous study are available, drawing comparisons with the power prior. The HPP allows for discounting based on the quality of individual level predictions, having the potential to provide efficiency gains (e.g., lower MSE) where predictions are incompatible with the data. An efficient Markov chain Monte Carlo algorithm is developed. Applications illustrate that inferences under the HPP are more robust to prior-data conflict compared to selected non-hierarchical priors.
\end{abstract}

\keywords{hierarchical models \and generalized linear models \and bayesian inference \and hyperprior}

\section{Introduction}
Exponential family models, which include distributions for binary, count, and continuous data, are among the most utilized models for statistical analysis. In many application areas, it is desirable to incorporate prior information in the analysis of data. In the Bayesian paradigm, such information may be incorporated through an informative prior distribution on the parameters of interest and, possibly, nuisance parameters. For example, in rare disease clinical trials, it can be difficult to establish treatment efficacy based solely on data from a prospective study. In ecological models, informative priors on the basis of similar studies, historical data, or expert opinion are often utilized (see, e.g., \citet{kuhnert_guide_2010} and \citet{banner_use_2020} and references therein). In applied econometric models, informative priors for generalized linear models (GLMs) have been, for example, utilized to predict unemployment rates \citep{grzenda_informative_2016} and to model the probability of credit default \citep{wang_improving_2018}.

When previous studies have been conducted, it is often desirable to construct a prior based on these data. Three popular priors that have been proposed for this setting are the power prior \citep{ibrahim_power_2000}, commensurate priors \citep{hobbs_commensurate_2012}, and meta-analytic-predictive (MAP) priors (e.g., \citet{schmidli_robust_2014}). The primary limitation of the former two priors is that they require access to an entire historical set in order to implement a joint prior for the regression coefficients that are correlated a priori. MAP priors are limited in that they have only been developed under the context where there is a single parameter of interest. In many data applications, it is desirable to elicit a joint prior for regression coefficients. For example, in clinical trials and observational studies, it is often of interest to determine whether an intervention is more/less efficacious for certain groups (i.e., effect modification). Another example arises when prediction is of paramount interest, such as in applied econometrics.

A multivariate normal prior on the parameters of interest and nuisance parameters may be utilized using summary statistics from publications for previously completed studies (e.g., as done in \citet{rodhouse_evidence_2019}). However, in regression models, this induces an a priori independent prior on the regression coefficients, which is primarily a function of constraints on the information source rather than any a priori belief supporting independence. In contrast, a posteriori, the posterior distribution for regression coefficients almost always exhibits dependence. 

Moreover, when a multivariate normal prior is elicited for a parameter on the basis of expert opinion, it may be difficult to translate the information provided to the regression parameter scale, and also to appropriately translate uncertainty. For example, in logistic regression models, a prior on a treatment effect would constitute a prior on the logarithm of the odds ratio. While one may transform an expert's opinion about the response probability to a log odds ratio, it is less clear how to elicit the variance on this prior. A second type of prior for exponential family models involves the elicitation of a prior prediction for the mean response (e.g., \citet{diaconis_conjugate_1979} for i.i.d. settings and \citet{chen_conjugate_2003} for regression settings). This type of prior may be more natural for expert elicitation than the previous, as the prior is elicited on the same scale as the data. However, these priors do not directly account for uncertainty in the prediction. Specifically, uncertainty may only be elicited through discounting of the informativeness of the prior.

In this paper, we develop a novel extension of the conjugate prior of \citet{chen_conjugate_2003}, where the prediction of the mean response is treated as random. Treatment of the prior prediction as random is intuitive because the prior prediction will typically made on the basis of summary statistics or expert opinion, both of which have some degree of uncertainty. We refer to the hierarchical prior as the hierarchical prediction prior (HPP). In regression models, the HPP induces a correlation structure on the regression coefficients a priori. The conjugate priors of \citet{diaconis_conjugate_1979} and \citet{chen_conjugate_2003} may be cast as special cases of the HPP for i.i.d. and regression models, respectively. Moreover, the HPP is quite flexible, enabling practitioners to elicit predictions on the mean response based on one or more covariates.

The posterior density under the HPP is robust to incompatible prior predictions for the mean response. In particular, we show for i.i.d. settings and the normal linear model that the posterior means of the predictions are between the predicted values based on the maximum likelihood estimator (MLE) and the elicited prior prediction. While a formal proof cannot be offered for all GLMs, the result held for a Poisson regression example. Moreover, under some limiting cases, we show for i.i.d. data settings and the normal linear model that the HPP is a conjugate prior for both the parameters of interest and the hierarchical parameters.

We compare the posterior distribution under the HPP for a logistic regression model using the data of \citet{finney_estimation_1947} against that of the conjugate prior of \citet{chen_conjugate_2003}, which we henceforth refer to as the CI prior. The results suggest that the posterior distribution of the regression coefficients under the CI prior is more sensitive to the prior prediction of the mean response than that for the HPP. We also illustrate how to utilize the HPP for settings in which a previous study was conducted, but only the point estimates and standard errors for the regression coefficients are available, comparing the posterior distribution of the parameter of interest under the HPP against that for the CI prior, an approximation to the asymptotic power prior \citep{ibrahim_power_2015}, and the power prior \citep{ibrahim_power_2000}, the latter for which the entire historical data set was utilized. When the historical data set is incompatible with the current data set (e.g., when the historical data provide evidence that treatment is beneficial and the current data suggest treatment is unbeneficial), the posterior of the treatment effect under the HPP consistently placed more mass at 0, while the posterior distributions for the treatment effect under the other priors suggested that treatment was all but certain to be efficacious.

In i.i.d. settings, samples from the posterior density under the HPP is straightforward utilizing any Markov-chain Monte Carlo (MCMC) algorithm. For regression settings excluding the normal linear model, sampling is more involved because there is no general closed-form of the normalizing constant for the CI prior. We utilize a fast and accurate Laplace approximation to the normalizing constant, which enables efficient sampling using MCMC methods.

The remainder of the article is organized as follows. 
In Section \ref{sec:priorelicit}, we review the conjugate prior of \citet{diaconis_conjugate_1979} and develop the HPP for independent and identically distributed (i.i.d.) data settings.
In Section \ref{sec:glm} we develop the HPP for generalized linear models (GLMs). 
In Section \ref{sec:previousstudies}, we illustrate how the HPP may be utilized for settings in which summary statistics are available (e.g., in a publication).
In Section \ref{sec:data}, we compare results from an analysis of the data of \citet{finney_estimation_1947} using the HPP and the CI prior for settings in which the prior is formulated on the basis of expert opinion.
In Section \ref{sec:sims}, we compare posterior quantities under the HPP with other priors based on data from a previously conducted study.
In Section \ref{sec:discussion}, we close with some discussion.

\section{The i.i.d. case}
\label{sec:priorelicit}
We begin by considering the independently and identically distributed (i.i.d.) data setting for exponential family models. We develop the HPP for the i.i.d. case and establish theoretical connections to the conjugate prior of \citet{diaconis_conjugate_1979}.

\subsection{Hyperprior motivation and construction}
Here, we discuss the motivation for the HPP and some of its properties. Suppose we observe $\{y_i, i = 1, \ldots, n \}$ with likelihood function in the exponential family, given by
\begin{align}
  f(\bm{y} | \theta, \phi) = \prod_{i=1}^n \exp\left\{ \frac{1}{a_i(\phi)} \left[ y_i \theta - b(\theta) \right] + c(y_i; \phi) \right\},
  \label{eq:iid_expfamdens}
\end{align}
where $a_i$ is a positive function and typically $a_i(\phi) = w_i \phi$, $\bm{y} = (y_1, \ldots, y_n)'$, $\theta \in \Theta$ is referred to as the canonical parameter, where $\Theta$ is the domain of $\theta$, and the functions $b$ and $c$ index the density or mass function. We assume $a_i(\phi) = \phi$ is known and fixed, and we may suppose without loss of generality that $\phi = 1$. \citet{diaconis_conjugate_1979} showed that each distribution in the exponential family admits a conjugate prior of the form
\begin{align}
    \DY(\theta | \lambda, m) 
      = \frac{1}{Z(\lambda, \lambda m)} \exp\left\{\lambda \left[ \theta m - b(\theta) \right] \right\},  \theta \in \Theta,
    \label{eq:iid_conjprior}
\end{align}
where 
$Z(\lambda, \lambda m) = \int_{\Theta} \exp\left\{\lambda \left[ \theta m - b(\theta) \right] \right\} d\theta$ 
is a normalizing constant, $\lambda$ is a precision parameter typically chosen so that $\lambda \in (0, n]$, and $m \in \dot{b}(\Theta)$ is a location parameter, where $\dot{b}$ is the first derivative of the function $b$ and $\dot{b}(\Theta)$ is the image of the set $\Theta$ under the function $\dot{b}$. Henceforth, we refer to the prior (\ref{eq:iid_conjprior}) simply as the ``DY prior.'' \citet{diaconis_conjugate_1979} showed that, under the DY prior, $E(y) = m$, that is, $m$ may be interpreted as a prior prediction (or ``guess'') for the mean of $y$. The hyperparameter $\lambda$ controls for the level of informativeness in the prior. Let $\bm{y} = (y_1, \ldots, y_n)'$. The posterior density utilizing the likelihood (\ref{eq:iid_expfamdens}) and prior (\ref{eq:iid_conjprior}) is given by
\begin{align*}
    p(\theta | \bm{y}, \lambda, m) &
    = \DY\left(\theta \left| n + \lambda, \frac{n\bar{y} + \lambda m}{n + \lambda} \right)\right.,
    %
\end{align*}
where $\bar{y} = n^{-1} \sum_{i=1}^n y_i$ is the sample mean. The primary disadvantage of the DY prior is that the posterior is sensitive to the prior prediction $m$ when $\lambda$ is large. For example, let $\mu = \dot{b}(\theta)$ denote the mean parameter of the distribution of the response variables. It can be shown that the posterior mean of $\mu$ under the DY prior is given by
$
E(\mu | \bm{y}, \lambda, m) = (n \bar{y} + \lambda m) / (n + \lambda),
$
i.e., the posterior mean of $\mu$ under the DY prior is a convex combination of the observed sample mean $\bar{y}$ and the prior prediction $m$, with higher values of $\lambda$ putting more weight on the value of $m$ in the posterior mean of $\mu$. Thus, if $m$ is an inaccurate prediction for $E(y)$, the posterior under the DY prior could give misleading results. 

Moreover, the hyperparameter $\lambda$ in the DY prior cannot simultaneously control the uncertainty surrounding $m$ and the level of borrowing from the prior. For example, in a Bernoulli model, suppose an expert believes with $95\%$ probability that the success probability is between 0.2 and 0.4. This may be accomplished by eliciting $m = 0.3$ and $\lambda = 78.8$ in the DY prior (\ref{eq:iid_conjprior}). This imposes a restriction on the sample size, namely, $n \ge 79$. Conversely, for fixed values of $n$, $\lambda$, and $m$, we cannot adjust the DY prior to account for uncertainty surrounding the value of $m$. 

Since $m$ is typically elicited on the basis of summary statistics or expert opinion, both of which are measured with uncertainty, it is natural to view $m$ as having a probability distribution. Thus, we propose to treat the hyperparameter $m$ in the DY prior as random. We refer to the joint prior of $(\theta, m)$ as the hierarchical prediction prior (HPP), and the prior on $m$ simply as the hyperprior.

We now develop the hyperprior. Let $\nu = \dot{b}(m) \in \Theta$. We elicit
\begin{align}
    \pi(\nu | \lambda_0, \mu_0) \propto \exp\left\{ \lambda_0[\nu \mu_0 - b(\nu)] \right\}, \nu \in \Theta,
    \label{eq:iid_nuprior}
\end{align}
which we may write in terms of the shape parameter $m$ as
\begin{align}
    \HPP(m | \lambda_0, \mu_0) \propto \exp\left\{ \lambda_0[ \dot{b}(m) \mu_0 - b(\dot{b}(m))] \right\} \frac{1}{v(m)}, m \in \dot{b}(\Theta),
    \label{eq:iid_mprior}
\end{align}
where $v(m) = \ddot{b} \circ \dot{b}^{-1}(m)$ is the variance function associated with the exponential family model (\ref{eq:iid_expfamdens}). For each exponential family model, the hyperprior is a recognizable density (e.g., a beta density for binomial models, a gamma density for Poisson models, an inverse-gamma density for gamma models, and a normal density for normal models). The hyperparameter $\lambda_0$ is a precision parameter controlling for the level of certainty surrounding the prior prediction.

The HPP is obtained by combining the DY prior (\ref{eq:iid_conjprior}) and the hyperprior (\ref{eq:iid_mprior}), giving
\begin{align}
    \HPP(\theta, m | \lambda, \lambda_0 \mu_0) = \CI(\theta | \lambda, m) \HPP(m | \lambda_0, \mu_0), \theta \in \Theta, m \in \dot{b}(\Theta).
    \label{eq:HPP}
\end{align}
The HPP is thus the CI prior conditional on $m$ multiplied by a density over $m$. It can be shown that the prior mean of $m$ in the hyperprior is $E(m) = \mu_0$. Thus,
$$
  E(y) = E_{\theta}[E(y|\theta)] = E_{\theta}[\dot{b}(\theta)] = E_{m}[ E_{\theta | m}(\dot{b}(\theta)) ] = E(m) = \mu_0,
$$
so that, similar to the DY prior, $\mu_0$ may be interpreted as a prior prediction for $E(y)$. However, unlike the DY prior, the HPP allows practitioners to directly elicit uncertainty surrounding the prior prediction $\mu_0$ for any fixed level of $\lambda$.

Note that as $\lambda_0 \to \infty$, $\HPP(m | \lambda_0, \mu_0)$ converges to a point mass at $\mu_0$, so that the HPP converges to the CI prior as $\lambda_0 \to \infty$. The marginal prior for $\theta$ under the HPP is given by
$$
\HPP(\theta | \lambda, \lambda_0, \mu_0) = \int \CI(\theta | \lambda, m) \HPP(m | \lambda_0, \mu_0) dm,
$$
i.e., the prior for $\theta$ is the CI prior conditional on $m$ averaged over a distribution over $m$. In the Bernoulli example above, we may elicit $\mu_0 = 0.30$ and $\lambda_0 = 78.8$, so that, a priori, $P(0.2 \le m \le 0.4) = 0.95$ for any value of $\lambda$. In general, unlike the DY prior, the HPP allows practitioners to elicit their uncertainty surrounding the prior prediction while fixing the amount of information borrowed from the prior.

Combining (\ref{eq:iid_expfamdens}), (\ref{eq:iid_conjprior}), and (\ref{eq:iid_mprior}) yields the joint posterior density
\begin{align}
    p(\theta, m | \bm{y}, \lambda, \lambda_0, \mu_0) 
      &\propto \DY\left(\theta | n + \lambda, \frac{n\bar{y} + \lambda m}{n + \lambda}\right) \frac{Z(n + \lambda, n\bar{y} + \lambda m)}{Z(\lambda, \lambda m)} \HPP(m | \lambda_0, \mu_0), \notag \\
    &\propto \DY\left(\theta | n + \lambda, \frac{n\bar{y} + \lambda m}{n + \lambda}\right)  p(m | \bm{y}, \lambda, \lambda_0, \mu_0).
    \label{eq:iid_jointpost}
\end{align}
The relationship (\ref{eq:iid_jointpost}) indicates that the joint posterior density may be expressed as the posterior using DY prior conditional on $m$ multiplied by the posterior density of $m$. Note that as $\lambda_0 \to \infty$, $p(m | \bm{y}, \lambda, \lambda_0, \mu_0)$ is simply a point mass at $m = \mu_0$, and the posterior distribution of $\theta$ thus converges to the posterior under the DY prior.

We may write the marginal posterior distribution of $\theta$ as
\begin{align}
    p(\theta | \bm{y}, \lambda, \lambda_0, \mu_0) \propto \int \DY\left(\theta | n + \lambda, \frac{n\bar{y} + \lambda m}{n + \lambda}\right) p(m | \bm{y}, \lambda, \lambda_0, \mu_0) dm
    .
    \label{eq:iid_theta_margpost}
\end{align}
The relationship (\ref{eq:iid_theta_margpost}) indicates that the marginal posterior density of $\theta$ under the HPP may be interpreted as the posterior utilizing the DY prior with a fixed value of $m$ averaged over the posterior distribution of $m$. In Section \ref{subsec:limit} we prove that the posterior mean of $m$ is between $\bar{y}$ and $\mu_0$ as $\lambda_0 \to \infty$. Simulation results suggest that the result holds in general. In effect, the marginal posterior distribution of $\theta$ depends on the data $\bm{y}$ in two ways: through the shape parameter in the posterior of the DY prior conditional on $m$, and in the marginal posterior distribution for $m$. Conversely, the posterior density of the DY prior treats $m = \mu_0$ as fixed. If $\mu_0$ is a poor prediction for $E(y)$, then inference on the posterior distribution of $\theta$ could provide misleading results.

\subsection{Limiting posterior distributions}
\label{subsec:limit}
As previously stated, the relationship (\ref{eq:iid_theta_margpost}) indicates that the marginal posterior distribution of $\theta$ under the HPP may be arrived at by taking posterior under the DY prior conditional on $m$ averaged over the posterior distribution of $m$, which depends on both the prior prediction $\mu_0$ as well as the observed sample mean $\bar{y}$. We may write the posterior distribution of $m$ as
\begin{align}
    p(m | \bm{y}, \lambda, \lambda_0, \mu_0) \propto \frac{Z(\lambda + n, \lambda m + n \bar{y})}{Z(\lambda, \lambda m)} \HPP(m | \lambda_0, \mu_0),
    \label{eq:iid_mpost}
\end{align}
where $Z(a, c) = \int_{\Theta} \exp\{ a [\theta (c / a) - b(\theta)] \}d\theta$ is the normalizing constant of the DY prior with precision parameter $c$ and shape parameter $a$ and $\HPP(m | \lambda_0, \mu_0)$ is defined in (\ref{eq:iid_mprior}). As $\lambda \to \infty$, the hyperprior becomes a conjugate density, whose posterior mean is a convex combination of the sample mean and the prior prediction. We state this formally in the following theorem.
\\

\noindent \textbf{Theorem 2.1}: Let $y_1, \ldots, y_n$ be observations from the exponential family (\ref{eq:iid_expfamdens}). Let the prior for $(\theta, m)$ be given by the HPP (\ref{eq:HPP}). Then
$$
    p(m | \bm{y}, \lambda, \lambda_0, \mu_0) \to \HPP\left( m \left| n + \lambda_0, \frac{n\bar{y} + \lambda_0 \mu_0}{n + \lambda_0} \right) \right. \text{ as } \lambda \to \infty.
$$
In words, Theorem 2.1 states that as $\lambda \to \infty$, the marginal prior of $m$ under the HPP is conjugate density. Hence, even as $\lambda \to \infty$, the posterior distribution of $\bm{m}$ is non-degenerate. Because the marginal prior for $m$ in the HPP is conjugate in the limit, Theorem 2.1 provides an approximation for the posterior of $m$ for large values of $\lambda$ in terms of recognizable densities (e.g., a gamma distribution for Poisson models).

An important consequence of Theorem 2.1 is
\begin{align*}
  E(m | \bm{y}, \lambda, \lambda_0, \mu_0) \to \frac{n \bar{y} + \lambda_0 \mu_0}{n + \lambda_0} \text{ as } \lambda \to \infty,
  %
\end{align*}
so that the posterior mean of $m$ converges to a convex combination of the observed sample mean, $\bar{y}$, and the elicited prior mean, $\mu_0$. This is closely related to, but quite different than, the posterior distribution of $\mu$ using the DY prior. The parameter $\mu$ is a model parameter, while the hierarchical parameter $m$ is a prior prediction for $\mu$. We prove Theorem 2.1 for the Poisson, Bernoulli, gamma, and normal models in Section 1 of the Supplementary Materials.

\noindent \textbf{Corollary 2.1} Given the same setup as Theorem 2.1, 
$$\lim_{\lambda \to \infty} p(\mu | \bm{y}, \lambda, \lambda_0, \mu_0) \to \HPP\left(\mu \left| n + \lambda_0, \frac{n\bar{y} + \lambda_0 \mu_0}{n + \lambda_0} \right) \right. .$$ 

\noindent \textit{Proof of Corollary 2.1}: Note that
$$
p(\mu | \bm{y}, \lambda, \lambda_0, \mu_0) = \int p(\mu | m, \bm{y}, \lambda, \lambda_0, \mu_0) p(m | \bm{y}, \lambda, \lambda_0, \mu_0) dm.
$$
As $\lambda \to \infty$, $p(\mu | m, \bm{y}, \lambda, \lambda_0, \mu_0)$ converges to a point mass at $m$ and $p(m | \bm{y}, \lambda, \lambda_0, \mu_0)$ converges to $\HPP(m | n + \lambda_0, (n\bar{y} + \lambda_0 \mu_0) / (n + \lambda_0))$ by Theorem 2.1. Hence, the result follows.

In words, Corollary 2.1 states that, as $\lambda \to \infty$, the posterior distribution for the mean parameter, $\mu$, under the HPP converges to the posterior density under the DY prior with precision parameter $n + \lambda_0$ and mean parameter $(n \bar{y} + \lambda_0 \mu_0) / (n + \lambda_0)$, which is precisely the same posterior distribution obtained by eliciting $\lambda = \lambda_0$ and $m = \mu_0$ in the DY prior (\ref{eq:iid_conjprior}). When $\lambda \le n$ (as is typically the case), the HPP cannot have more influence on the posterior than the likelihood. This fact and Corollary 2.1 imply that if $\lambda \to \infty$ and $\lambda_0 \le n$ or $\lambda_0 \to \infty$ and $\lambda \le n$, the posterior distribution of $\mu$ cannot depend more on the prior than the data. 

\textbf{Remark 2.1}: By transforming the mean parameter, $\mu$, to the canonical parameter $\theta$, it is easy to see from Corollary 2.1 that $p(\theta | \bm{y}, \lambda, \lambda_0, \mu_0) \to \DY(\theta | n + \lambda_0, (n\bar{y} + \lambda_0 \mu_0)/(n + \lambda_0) )$ as $\lambda \to \infty$, therefore, the posterior density of $\theta$ under the HPP converges to that under the DY prior, with shape parameter equal to a convex combination of the sample mean and prior prediction.

\textbf{Remark 2.2}: For the i.i.d. normal case, the HPP induces a conjugate prior on $\theta = \mu$ for finite $\lambda$. The difference between the DY prior and the HPP for the normal case is that the prior variance of $\mu$ under the HPP is larger than the DY prior for $\lambda_0 < \infty$.

\section{The regression case}
\label{sec:glm}
In this section, we develop the HPP and illustrate its properties for GLMs. We derive the posterior distribution of the regression coefficients for the normal linear model under the HPP, relating it to the posterior under the CI and multivariate normal priors. Furthermore, we discuss the posterior distribution of $\bm{m}$ and how possible efficiency gains may be made achieved even when some of the components of $\bm{\mu}_0$ are inaccurate. Finally, we discuss how to implement the HPP computationally to obtain posterior samples.

\subsection{The HPP for GLMs}
Throughout this section, suppose we observe $\{(y_i, \bm{x}_i), i = 1, \ldots, n\}$, where $y_i$ is a response variable and $\bm{x}_i$ is a $p \times 1$ vector of covariates associated with subject $i$, which may include an intercept term. Suppose $E(y_i) = \mu_i$, where $g(\mu_i) = \bm{x}_i' \bm{\beta}$, where $\bm{\beta}$ is a $p \times 1$ vector of regression coefficients. The function $g$ is referred to as the $\mu$-link function. The likelihood function of $\bm{y} = (y_1, \ldots, y_n)'$ may be written as
\begin{align}
    f(\bm{y} | \phi, \bm{\beta}, \bm{X}) = \prod_{i=1}^n \exp\left\{ \frac{1}{a_i(\phi)} \left[ y_i \theta(\bm{x}_i'\bm{\beta}) - b(\theta(\bm{x}_i'\bm{\beta})) \right] + c(y_i, \phi) \right\},
    \label{eq:glm_jointdensity}
\end{align}
where $\theta(x) = \dot{b}^{-1} \circ g^{-1}(x)$ is referred to as the $\theta$-link function and the functions $b$ and $c$ index the density or mass function. We assume $a_i(\phi) = \phi$ for $i = 1, \ldots, n$ is known and fixed, and we may suppose without loss of generality that $\phi = 1$. \citet{chen_conjugate_2003} showed that the likelihood (\ref{eq:glm_jointdensity}) admits a conjugate prior of the form
\begin{align}
    \CI(\bm{\beta} | \lambda, \bm{m}) = \frac{1}{Z(\lambda, \lambda \bm{m})} \exp\left\{ \lambda \left[ \bm{m}'\theta(\bm{X\beta}) - \bm{J}'b(\theta(\bm{X\beta})) \right] \right\}, \bm{\beta} \in \mathbb{R}^p,
    \label{eq:glm_conjprior}
\end{align}
where $Z(\lambda, \lambda \bm{m}) = \int_{\mathbb{R}^p} \exp\left\{ \lambda \left[ \bm{m}'\theta(\bm{X\beta}) - \bm{J}'b(\theta(\bm{X\beta})) \right] \right\} d\bm{\beta}$ is the normalizing constant, which has no closed-form expression in general, $\lambda > 0$ is a precision parameter controlling for the informativeness of the prior, $\bm{m}$ is a $n$-dimensional shape parameter that may be interpreted as a prior prediction for $E(\bm{y})$, and the function $b$ is taken componentwise, i.e., $b(\bm{\eta}) = (b(\eta_1), \ldots, b(\eta_n))'$ for $\eta \in \mathbb{R}^n$. We refer to the prior (\ref{eq:glm_conjprior}) as the ``CI prior.'' When $\bm{m}$ is fixed, \citet{chen_conjugate_2003} showed that the posterior is
\begin{align}
    p(\bm{\beta} | \lambda, \bm{m}) &= \CI\left(\bm{\beta} \left| 1 + \lambda, \frac{\bm{y} + \lambda \bm{m}}{1 + \lambda} \right) \right. .
    \label{eq:glm_conjpost}
\end{align}
Typically, $\lambda \in (0, 1]$ so that the effective sample size contributed by the prior does not exceed that of the data. Note that, similar to the i.i.d. case, the posterior distribution of $\bm{\beta}$ for fixed $\bm{m}$ has a shape parameter that is a convex combination of the observed data $\bm{y}$ and prior prediction $\bm{m}$. Thus, it is clear that the posterior (\ref{eq:glm_conjpost}) is sensitive to the choice of $\bm{m}$.

One of the challenges encountered when using the CI prior (\ref{eq:glm_conjprior}) is that it is difficult to quantify uncertainty surrounding the value of $\bm{m}$. That is, the CI prior treats $\bm{m}$ as observed data. While the precision hyperparameter $\lambda$ controls for the degree of influence of the prior on the posterior, it does not explicitly reflect uncertainty in the prior prediction. For example, similar to the i.i.d. case, $\lambda$ cannot simultaneously control for the uncertainty surrounding $\bm{m}$ as well as the discounting of the CI prior. Moreover, elicitation of $\bm{m}$ is typically based on expert opinion or summary statistics, and, thus, it is natural to view $\bm{m}$ as having a probability distribution.

To this end, we now derive the hyperprior. Let $\bm{\nu} = \dot{b}(\bm{m})$, where the function $\dot{b}$ is taken component-wise. We may elicit
\begin{align}
    \pi(\bm{\nu} | \lambda_0, \bm{\mu}_0) \propto \exp\left\{ \lambda_0 \left[ \bm{\nu}'\bm{\mu}_0 - \bm{J}'b(\bm{\nu}) \right] \right\} = \prod_{i=1}^n \exp\left\{ \lambda_0 \left[ \nu_i \mu_{0i} - b(\nu_i) \right] \right\},
    \bm{\nu} \in \Theta^{n}
    \label{eq:glm_nuprior},
\end{align}
where $\bm{\nu} = \dot{b}(\bm{m})$, $\lambda_0 > 0$ is a precision parameter, and $\bm{\mu}_0 \in [\dot{b}(\Theta)]^n$ is a $n$-dimensional vector giving the prior prediction for $E(\bm{y})$, which may depend on covariates. In general, one may utilize a separate precision parameter for each component of $\bm{\nu}$, but we will proceed with a common precision parameter for notational convenience (allowing each component to have its own precision is discussed in Section \ref{sec:previousstudies}). Note that (\ref{eq:glm_nuprior}) is a product of $n$ independent priors of the form (\ref{eq:iid_nuprior}). Thus the prior for regression settings (\ref{eq:glm_nuprior}) can be thought of as a multivariate generalization of that for i.i.d. settings (\ref{eq:iid_nuprior}). For example, if $\bm{\nu} = \nu$ is a scalar, then (\ref{eq:glm_nuprior}) is equal to (\ref{eq:iid_nuprior}).

Using (\ref{eq:glm_nuprior}), the hyperprior on the mean scale is obtained using the transformation $\bm{m} = \dot{b}^{-1}(\bm{\nu})$, i.e., 
\begin{align}
    \HPP(\bm{m} | \lambda_0, \bm{\mu}_0) \propto \prod_{i=1}^n \exp\left\{ \lambda_0 \left[ \dot{b}^{-1}(m_i)'\mu_{0i} - b(\dot{b}^{-1}(m_i)) \right] \right\}    \frac{1}{v(m_i)}, \bm{m} \in [\dot{b}(\Theta)]^n,
    \label{eq:glm_mprior}
\end{align}
where $v(x) = \ddot{b} \circ \dot{b}^{-1}(x)$ is the variance function of the family. Note that the hyperprior (\ref{eq:glm_mprior}) is a product of $n$ independent densities, each having the same form as the i.i.d. case (\ref{eq:iid_mprior}), where each component of $\bm{m}$ has its own mean. The hyperprior is thus a product of $n$ independent recognizable densities (e.g., beta for binomial models, gamma for Poisson models, inverse-gamma for gamma models, and normal for normal models). An attractive feature of the HPP is that, if covariates and regression coefficients are ignored, the HPP is a conjugate prior for each component of $\bm{y}$. For example, if $ \bm{y} = (y_1, \ldots, y_n)'$ is a collection of Bernoulli random variables each with mean $\mu_{i}$, $\{ m_i, i = 1, \ldots, n \}$ is a priori a collection of $n$ independent beta random variables with mean $\mu_{0i}$ and dispersion parameter $\lambda_0$.

In some cases, it is desirable to elicit a prior on $\bm{m}$ based on (possibly a subset of) covariates, which we view as fixed. For example, we may decompose $\bm{\beta} = (\bm{\alpha}', \bm{\gamma}')'$ and $\bm{x}_i = (\bm{x}_{1i}', \bm{x}_{2i}')'$. If $\bm{\alpha}_0$ is a prediction for $\bm{\alpha}$, we may take $\mu_{0i} = g^{-1}(\bm{x}_{1i}'\bm{\alpha}_0)$, where $g$ is the $\mu$-link function. In this case, we may write (\ref{eq:glm_mprior}) as
\begin{align}
    \HPP(\bm{m} | \lambda_0, \bm{\mu}_0) \propto \exp\left\{ \lambda_0 \left[ \dot{b}^{-1}(\bm{m})' g^{-1}(\bm{X}_1\bm{\alpha}_0) - \bm{J}' b(\dot{b}^{-1}(\bm{m})) \right] \right\}.
    \label{eq:glm_mprior_reg}
\end{align}
For example, in a Poisson regression model with canonical link function, we have $b(x) = \exp\{x\}$ and $g(\mu) = \exp\{\mu\}$, so that we may write the hyperprior (\ref{eq:glm_mprior_reg}) as
$$
    \HPP(\bm{m} | \lambda_0, \bm{\alpha}_0 ) \propto \exp\left\{ \lambda_0 \left[ \log(\bm{m})' \exp\{ \bm{X}_1\bm{\alpha}_0 \} - \bm{J}'\bm{m} \right] \right\}.
$$
Thus, the hyperprior is proportional to the likelihood of a gamma GLM with known inverse dispersion parameter $\lambda_0$, known regression coefficients $\bm{\alpha}_0$, and $\mu$-link function $g(\mu) = \log(\mu)$. Hence, the hyperprior (\ref{eq:glm_mprior_reg}) may be thought of as a random process for predicting the mean response based on a prediction $\bm{\alpha}_0$ based on covariates $\bm{X}_1$.

The HPP for regression models is obtained by combining the CI prior (\ref{eq:glm_conjprior}) and the hyperprior (\ref{eq:glm_mprior_reg}), giving
\begin{align}
    \HPP(\bm{\beta}, \bm{m} | \lambda, \lambda_0, \bm{\mu}_0) &= \CI(\bm{\beta} | \lambda, \bm{m}) \HPP(\bm{m} | \lambda_0, \bm{\mu}_0).
    \label{eq:hpp_reg}
\end{align}
Thus, the marginal prior for the regression coefficients is
$$
\HPP(\bm{\beta} | \lambda, \lambda_0, \mu_0) = \int \CI(\bm{\beta} | \lambda, \bm{m}) \HPP(\bm{m} | \lambda_0, \bm{\mu}_0) d\bm{m}.
$$
The marginal prior of $\bm{\beta}$ under the HPP is thus the CI prior conditional on $\bm{m}$ averaged over a distribution on $\bm{m}$. As uncertainty surrounding the prior prediction decreases, i.e., for larger values of $\lambda_0$, the marginal prior of $\bm{\beta}$ under the HPP will become more similar to the CI prior. In particular, as $\lambda_0 \to \infty$, the HPP and the CI priors coincide. 

Using the HPP (\ref{eq:hpp_reg}), the joint posterior density may be written as
\begin{align}
    p(\bm{\beta}, \bm{m} | \bm{y}, \lambda, \lambda_0, \bm{\mu}_0) 
    &\propto 
    \CI\left( \bm{\beta} \left| 1 + \lambda, \frac{\bm{y} + \lambda \bm{m}}{1 + \lambda} \right) \right.
    \frac{Z(1 + \lambda, \bm{y} + \lambda \bm{m})}{Z(\lambda, \lambda \bm{m})}\HPP(\bm{m} | \lambda_0, \bm{\mu}_0), \notag \\
    &\propto p(\bm{\beta} | \bm{y}, \lambda, \bm{m}) p(\bm{m} | \bm{y}, \lambda, \lambda_0, \mu_0).
    \label{eq:glm_hierpost}
\end{align}
Thus, the joint posterior density may be expressed as the product of the posterior under the CI prior with shape parameter $\frac{\bm{y} + \lambda \bm{m}}{1 + \lambda}$ and a density over $\bm{m}$, which depends on the observed data $\bm{y}$ and the prior prediction $\bm{\mu}_0$. For intercept-only models, i.e., under the special case where $\bm{\beta} = \beta_0$ and $\bm{X} = \bm{J}$, $\bm{m} = m$ is a scalar and it can be shown that
$\CI(\bm{\beta} | 1 + \lambda, (y + \lambda \bm{m}) / (1 + \lambda)) = \DY(\beta_0 | n + \lambda, (n\bar{y} + \lambda m) / (n + \lambda))$. Thus, the joint posterior (\ref{eq:glm_hierpost}) may be viewed as a multivariate extension of the joint posterior for the i.i.d. setting, given by (\ref{eq:iid_jointpost}).

The CI prior (\ref{eq:glm_conjprior}) is a member of a special class of priors called $g$-priors \citep{zellner_assessing_1986}. The case where the hyperparameter $\lambda$ is treated as random has been considered by \citet{sabanes_bove_hyper-g_2011} and further expanded upon in \citep{held_approximate_2015}. When $\lambda$ is treated as random and $\bm{m}$ is treated as fixed, the posterior distribution of $\lambda$ reflects how compatible the prior prediction $\bm{m}$ is with the observed data. For example, if many of the components of $\bm{m}$ differ from the observed data, the posterior distribution of $\lambda$ may be concentrated near 0, so that the prior has little effect in the posterior. By contrast, the HPP treats the prior prediction $\bm{m}$ as random and $\lambda$ as fixed, and, thus, the posterior distribution of the regression coefficients depends on the posterior distribution of $\bm{m}$, which depends on both the prior prediction $\bm{\mu}_0$ and the observed data. In Section \ref{subsec:mpost}, we discuss in more detail the important role that the posterior distribution of $\bm{m}$ has in the posterior distribution of the regression coefficients.

The HPP for GLMs is similar to a hierarchical conditional means prior (CMP) for GLMs \citep{bedrick_new_1996}. For the CMP, $p$ \emph{potential} response and covariate pairs $(\tilde{y}_i, \tilde{\bm{x}}_i)$ are elicited, where $p$ is the number of regression coefficients. The $\tilde{y}_i$'s may be interpreted as a prior prediction for the mean response based on covariate $\tilde{x}_i$. A prior inducing an a priori correlation structure on the regression coefficients may be obtained by treating each $\tilde{y}_i, i = 1, \ldots, p$ as random (for example, by utilizing the hyperprior of the HPP). However, it may be difficult to justify a choice for the $p$ potential covariate vectors $\tilde{x}_i$. By contrast, specification of $\bm{\mu}_0$ is straightforward because it is simply a prior prediction of the mean response for each of the $n$ observations, potentially based on \emph{observed} (rather than potential) covariates.

\subsection{The linear model}
We now explore the posterior under the HPP for the normal linear model. Specifically, we show that the marginal prior for the regression coefficients under the HPP and the corresponding posterior density are multivariate normal. Furthermore, we draw comparisons of the HPP with the CI prior and a multivariate normal prior. We generalize Theorem 2.1 for the normal linear model, showing that the limiting distribution of the posterior distribution of $\bm{\beta}$ under the HPP as $\lambda \to \infty$ is the posterior density under the CI prior.

Suppose we observe data $D = \{(y_i, \bm{x}_i), i = 1, \ldots, n\}$ with likelihood function
$$
L(\bm{\beta}, \tau | D) \propto \tau^{-n/2} \exp\left\{ -\frac{\tau}{2} (\bm{y} - \bm{X\beta})'(\bm{y} - \bm{X\beta}) \right\},
$$
where $\tau = 1 / \sigma^2$ is a precision parameter, $\bm{\beta}$ is a $p$-dimensional vector of regression coefficients, $\bm{y} = (y_1, \ldots, y_n)'$ is a $n$-dimensional response vector, and $\bm{X} = (\bm{x}_1, \ldots, \bm{x}_n)'$ is a $n \times p$ matrix (which may include an intercept term). We assume that $\tau$ is known and fixed, and we may assume without loss of generality that $\tau = 1$.

For the normal linear model, the CI prior is given by
\begin{align}
    \CI(\bm{\beta} | \lambda, \bm{m}) \propto \exp\left\{ -\frac{\lambda}{2} \left[ \bm{\beta}'\bm{X}'\bm{X} \bm{\beta} - 2 \bm{\beta}' \bm{X}'\bm{X} \hat{\bm{\beta}}_{\bm{m}} \right] \right\},
    \label{eq:lm_ciprior}
\end{align}
where $\hat{\bm{\beta}}_{\bm{m}} = (\bm{X}'\bm{X})^{-1} \bm{X}'\bm{m}$. Note that, for a fixed value of $\bm{m}$, $\hat{\bm{\beta}}_{\bm{m}}$ is the MLE for a normal linear model with vector of response variables $\bm{m}$ and design matrix $\bm{X}$. The normalizing constant $Z(\lambda, \lambda\bm{m})$ of the CI prior is given by
\begin{align}
    Z(\lambda, \lambda\bm{m}) 
    &= (2\pi)^{p/2} \lvert \lambda^{-1} (\bm{X}'\bm{X})^{-1} \rvert^{1/2} \exp\left\{ \frac{\lambda}{2} \left[ \hat{\bm{\beta}}_{\bm{m}}' \bm{X}'\bm{X} \hat{\bm{\beta}}_{\bm{m}}\right] \right\}.
    \label{eq:lm_normconst}
\end{align}
It follows from (\ref{eq:lm_ciprior}) and (\ref{eq:lm_normconst}) that the the CI prior on $\bm{\beta}$ for a fixed value of $\bm{m}$ is multivariate normal, i.e.,
$$
    \CI(\bm{\beta} | \lambda, \bm{m}) = \phi_p(\bm{\beta} | \hat{\bm{\beta}}_{\bm{m}}, \lambda^{-1}(\bm{X}'\bm{X})^{-1}),
$$
where $\phi_p( \bm{x} | \bm{a}, \bm{C}) = (2\pi)^{-p/2} \lvert \bm{C} \rvert^{-1/2} \exp\left\{ (\bm{x} - \bm{a})' \bm{C}^{-1} (\bm{x} - \bm{a}) \right\}$ is the $p$-dimensional multivariate normal density with mean $\bm{a}$ and positive definite covariance matrix $\bm{C}$. The hyperprior (\ref{eq:glm_mprior}) is also a multivariate normal density. That is,
\begin{align}
    \HPP(\bm{m} | \lambda_0, \bm{\mu}_0) \propto \phi_n(\bm{m} | \bm{\mu}_0, \lambda_0^{-1} \bm{I}_n),
    \label{eq:lm_hpp}
\end{align}
where $\bm{I}_n$ is a $n\times n$ identity matrix. Hence, a priori, $\bm{m} = (m_1, \ldots, m_n)'$ is a vector of $n$ independent normal random variables whose $i^{th}$ component has mean $\mu_{0i}$ and variance $\lambda_0^{-1}$.

We now derive the joint prior distribution of $\bm{\beta}$ and $\bm{m}$ for the normal linear model. We show that the joint posterior distribution of $(\bm{\beta}, \bm{m})$ is multivariate normal. We further discuss how the values of $\lambda_0$ and $\bm{\mu}_0$ affect the prior on $\bm{\beta}$. Note that the marginal mean and covariance matrix of $\bm{\beta}$ based on the HPP are given by
$$
E(\bm{\beta}) = E[E(\bm{\beta} | \bm{m})] = E[(\bm{X}'\bm{X})^{-1} \bm{X}' \bm{m}] = (\bm{X}'\bm{X})^{-1}\bm{X}'\bm{\mu}_0
$$
and
$$
\Cov(\bm{\beta}) = E\left[ \Cov(\bm{\beta} | \bm{m}) \right] + \Cov[E(\bm{\beta} | \bm{m})] 
                 = (\lambda^{-1} + \lambda_0^{-1})(\bm{X}'\bm{X})^{-1},
$$
so that the joint prior of $(\bm{\beta}, \bm{m})$ is multivariate normal, i.e.,
\begin{align}
    \HPP(\bm{\beta}, \bm{m}) = \phi_{n + p}\left( 
      \begin{pmatrix} \hat{\bm{\beta}}_{\bm{\mu}_0} \\ \bm{\mu}_0 \end{pmatrix},
      \begin{pmatrix}
      (\lambda^{-1} + \lambda_0^{-1}) (\bm{X}'\bm{X})^{-1} & \lambda_0^{-1} (\bm{X}'\bm{X})^{-1} \bm{X}' \\
      \lambda_0^{-1} \bm{X}' (\bm{X}'\bm{X})^{-1}          & \lambda_0^{-1} \bm{I}_n
      \end{pmatrix}
      \right),
      \label{eq:lm_jointprior}
\end{align}
where $\bm{\hat{\beta}}_{\bm{\mu}_0} = (\bm{X}'\bm{X})^{-1}\bm{X}'\bm{\mu}_0$ is the MLE for a normal linear model treating $\bm{\mu}_0$ as data. Hence, for the normal linear model, the prior mean of $\bm{\beta}$ induced by the HPP is precisely the same as the CI prior, treating $\bm{m} = \bm{\mu}_0$ as fixed, and the prior variances of the components of $\bm{\beta}$ under the HPP are larger than the CI prior. Moreover, (\ref{eq:lm_jointprior}) indicates that larger values of $\lambda_0$ shrink the correlation between $\bm{\beta}$ and $\bm{m}$ toward 0.

Using (\ref{eq:lm_jointprior}), we may write the marginal prior of $\bm{\beta}$ as
\begin{align}
    \HPP(\bm{\beta} | \lambda, \lambda_0, \bm{\mu}_0) = \int \HPP(\bm{\beta}, \bm{m} | \lambda, \lambda_0, \bm{\mu}_0)) d\bm{m} = \phi_p(\bm{\beta} | \hat{\bm{\beta}}_{\bm{\mu}_0}, (\lambda_{\text{H}} (\bm{X}'\bm{X}))^{-1}),
    \label{eq:lm_margprior_beta}
\end{align}
where $\lambda_{\text{H}} = (\lambda_0 \lambda) / (\lambda_0 + \lambda)$. Note that the marginal prior (\ref{eq:lm_margprior_beta}) has the form of the CI prior with $\bm{m} = \bm{\mu}_0$ and $\lambda = \lambda_{\text{H}}$. The connection between the marginal prior of $\bm{\beta}$ based on the HPP and the CI prior holds only when each component of $\bm{m}$ in the HPP has the same precision parameter $\lambda_0$. While it is permissible to allow each component of $\bm{m}$ to have its own degree of uncertainty in the HPP, the CI prior cannot simultaneously reflect uncertainty about $\bm{m}$ and control the level of discounting. Note also that the marginal posterior (\ref{eq:lm_margprior_beta}) may be obtained by eliciting a multivariate normal prior on $\bm{\beta}$ with mean $\hat{\bm{\beta}}_{\bm{\mu}_0}$ and covariance $(\lambda_{\text{H}} \bm{X}'\bm{X})^{-1}$ directly.


We now derive the posterior density of the regression coefficients for the normal linear model. Since, as stated above, the marginal prior of $\bm{\beta}$ is $\HPP(\bm{\beta} | \lambda, \lambda_0, \bm{\mu}_0) \propto \CI(\bm{\beta} | \lambda_{\text{H}}, \bm{\mu}_0)$, we have via the conjugacy of the CI prior that
\begin{align}
    p(\bm{\beta} | \bm{y}, \lambda, \lambda_0, \bm{\mu}_0) \propto \CI\left(\bm{\beta} | 1 + \lambda_{\text{H}}, \frac{\bm{y} + \lambda_{\text{H}} \bm{\mu}_0}{1 + \lambda_{\text{H}}}\right) \propto \phi_p(\bm{\beta} | \bm{\mu}_{\bm{\beta}}, (1 + \lambda_{\text{H}})^{-1} (\bm{X}'\bm{X})^{-1} ),
    \label{eq:glm_normal_post}
\end{align}
where
$$
  \bm{\mu}_{\bm{\beta}} = \frac{1}{1 + \lambda_{\text{H}}} \hat{\bm{\beta}} + \left(1 - \frac{1}{1 + \lambda_{\text{H}}} \right) \hat{\bm{\beta}}_{\bm{\mu}_0},
$$
where $\hat{\bm{\beta}} = (\bm{X}'\bm{X})^{-1} \bm{X}'\bm{y}$ is the MLE for $\bm{\beta}$. Hence, the marginal posterior distribution of $\bm{\beta}$ is a multivariate normal distribution with mean $\bm{\mu}_{\bm{\beta}}$ and covariance matrix $(1 + \lambda_{\text{H}})^{-1} (\bm{X}'\bm{X})^{-1}$. Note that $\bm{\mu}_{\bm{\beta}}$ is a convex combination of the MLE of $\bm{\beta}$ based on $\bm{y}$ and the ``MLE'' based on treating $\bm{\mu}_0$ as observed data. Since $\lambda_\text{H}$ increases as either $\lambda_0$ or $\lambda$ increases, larger values of $\lambda$ or $\lambda_0$ cause the posterior mean of $\bm{\beta}$ to be closer to $\bm{\mu}_0$ than smaller values. Note that when $\lambda \le 1$, $\lambda_{\text{H}} \le \lambda_0 / (1 + \lambda_0) < 1$. Hence, the posterior mean of $\bm{\beta}$ will always be closer to the MLE than $\hat{\bm{\beta}}_{\bm{\mu}_0}$ when $\lambda \le 1$.

When $\lambda_{\text{H}}$ is small, i.e., for small values of $\lambda$ or $\lambda_0$, the posterior covariance matrix of $\bm{\beta}$ becomes similar to that when using an improper uniform prior for the regression coefficients. This is intuitive, as small values of $\lambda$ induce less borrowing from the CI prior, and small values of $\lambda_0$ indicate substantial uncertainty surrounding the prior prediction $\bm{\mu}_0$.

An interesting limiting case is when $\lambda \to \infty$. Note that $\lim_{\lambda \to \infty} \lambda_{\text{H}} = \lambda_0$. Hence, it is clear from (\ref{eq:glm_normal_post}) that
\begin{align}
    p(\bm{\beta} | \bm{y}, \lambda, \lambda_0, \bm{\mu}_0) \to 
    \phi_p\left( \bm{\beta} \left| \frac{ \hat{\bm{\beta}} + \lambda_0 \hat{\bm{\beta}}_{\bm{\mu}_0} } { 1 + \lambda_0 }, \frac{(\bm{X}'\bm{X})^{-1}}{1 + \lambda_0} \right) \right. 
    \text{ as } \lambda \to \infty.
    \label{eq:glm_normal_post_limit}
\end{align}
Hence, as $\lambda \to \infty$, the posterior distribution of $\bm{\beta}$ is a multivariate normal distribution with posterior mean equal to a weighted average of $\hat{\bm{\beta}}$ and $\hat{\bm{\beta}}_{\bm{\mu}_0}$. The relationship (\ref{eq:glm_normal_post_limit}) indicates that, when $\lambda$ is allowed to unboundedly increase, the posterior distribution is non-degenerate, but depends heavily on the value of $\bm{\mu}_0$ when $\lambda_0$ is large. In practice, we typically restrict $\lambda \in (0, 1]$, so that the prior does not have more weight on the posterior than the data. However, the form of the CI prior and the relationship (\ref{eq:glm_normal_post_limit}) indicates that the posterior of $\bm{\beta}$ will depend more on the data than the prior prediction whenever $\lambda < 1$ or $\lambda_0 < 1$.

\subsection{The posterior distribution of m}
\label{subsec:mpost}
We now discuss the posterior distribution of $\bm{m}$. We will see that the posterior distribution of $\bm{m}$ has important implications on the posterior distribution of the regression coefficients. Namely, because the posterior distribution of the regression coefficients under the HPP is the posterior distribution of the CI prior averaged over the posterior of $\bm{m}$, both of which depend on the observed data, the HPP tends to put more weight on the observed data than the prior prediction in the posterior of the regression coefficients compared to the CI prior.

We may write the posterior distribution of $\bm{m}$ under the HPP as
\begin{align}
    p(\bm{m} | \bm{y}, \lambda, \lambda_0, \bm{\mu}_0) \propto \frac{Z(1 + \lambda, \bm{y} + \lambda \bm{m})}{Z(\lambda, \lambda \bm{m})} \HPP(\bm{m} | \lambda_0, \bm{\mu}_0),
    \label{eq:glm_mpost}
\end{align}
where $Z(a, \bm{c}) = \int \exp\left\{ a\left[(\bm{c}'/a) \theta(\bm{X\beta}) - b(\theta(\bm{X\beta})) \right] \right\}$ is the normalizing constant of the CI prior (\ref{eq:glm_conjprior}), which has no general closed-form solution. For the i.i.d. case, it was shown in Section \ref{subsec:limit} that, as $\lambda \to \infty$, the posterior mean of $m$ was between the prior prediction $\mu_0$ and the MLE $\hat{\mu} = \bar{y}$. Similarly, for GLMs, simulation studies suggest that the posterior mean of $\bm{m}$ is between $\hat{\bm{\mu}} = g^{-1}(\bm{X}\hat{\bm{\beta}})$ and $\bm{\mu}_0$, even for finite values of $\lambda$.

Note that we may write the posterior distribution of $\bm{\beta}$ under the HPP as
\begin{align}
    p(\bm{\beta} | \lambda, \lambda_0, \bm{\mu}_0) = \int \CI\left(\bm{\beta} | 1 + \lambda, \frac{\bm{y} + \lambda \bm{m}}{1 + \lambda} \right) p(\bm{m} | \lambda, \lambda_0, \bm{\mu}_0) d\bm{m}.
    \label{eq:glm_margpostbeta}
\end{align}
The relationship (\ref{eq:glm_margpostbeta}) indicates that the posterior distribution of $\bm{\beta}$ may be interpreted as the CI density with shape parameter $(\bm{y} + \lambda \bm{m}) / (1 + \lambda)$ averaged over the posterior distribution of $\bm{m}$, which depends on the data $\bm{y}$. Thus, the posterior distribution of $\bm{\beta}$ depends on the data in two ways: through the conditional posterior of the CI prior (treating $\bm{m}$ as known) and through the posterior of $\bm{m}$. 

When $\lambda$ is treated as random and $\bm{m}$ is treated as fixed (e.g., as done in \citet{sabanes_bove_hyper-g_2011}), the posterior distribution of $\lambda$ reflects how accurate the prior prediction is \emph{overall}. For example, if many of the prior predictions are inaccurate and some are highly accurate, the posterior distribution of $\lambda$ may be concentrated near 0, so that the prior has little influence on the posterior. Conversely, in the HPP, $\bm{m}$ is treated as random and $\lambda$ is fixed, so that the posterior always borrows from the HPP. The HPP discounts at the prediction level, so that it is possible for efficiency gains (e.g., lower mean square error) to be made even when some of the prior predictions are inaccurate.

For the normal linear model, it can be shown that the posterior distribution of $\bm{m}$ under the HPP is multivariate normal with mean
$
E(\bm{m} | \bm{y}) = \bm{\Lambda} \bm{\mu}_0 + (\bm{I}_n - \bm{\Lambda}) \hat{\bm{y}},
$
where $\bm{\Lambda} = \left(\lambda_0 \bm{I}_n + \frac{\lambda}{1 + \lambda} \bm{H} \right)^{-1} \lambda_0 \bm{I}_n$, $\bm{H} = \bm{X}(\bm{X}'\bm{X})^{-1}\bm{X}'$ is the orthogonal projection operator onto the space spanned by the columns of $\bm{X}$, and $\bm{\hat{y}} = \bm{X} \hat{\bm{\beta}}$ is the predicted values of $\bm{y}$, where $\hat{\bm{\beta}}$ is the MLE of $\bm{\beta}$. Details of the derivation are given in Section 3 of the Supplementary Materials. Hence, the posterior mean of $\bm{m}$ for the normal linear model is a convex combination of the prior prediction $\bm{\mu}_0$ and the predicted values $\hat{\bm{y}}$. Note that for larger values of $\lambda$ (i.e., when more information from the prior is borrowed) and for fixed values of $\lambda_0$, the posterior mean of $\bm{m}$ depends more on the predicted values than the prior prediction. This fact and the relationship (\ref{eq:glm_mpost}) illustrate the robustness of the HPP, namely, the posterior distribution of the regression coefficients is averaged over a distribution depending on the prior prediction $\bm{\mu}_0$, but highly modified by the data $\bm{y}$.

\subsection{Computational development}
\label{subsec:computation}
We now discuss how to obtain posterior samples under the HPP. We develop a Laplace approximation to the normalizing constant of the CI prior. We then propose an efficient Hamiltonian Monte Carlo algorithm to obtain posterior samples.

The joint posterior may be written as
\begin{align}
    p(\bm{\beta}, \bm{m} | \bm{y}, \lambda, \lambda_0, \mu_0) \propto \CI\left(\bm{\beta} \left| 1 + \lambda,  \frac{\bm{y} + \lambda \bm{m}}{1 + \lambda} \right) \right. \frac{\HPP(\bm{m} | \lambda_0, \bm{\mu}_0)}{Z(\lambda, \lambda \bm{m})}.
    \label{eq:glm_jointpost}
\end{align}
While posterior inference in the i.i.d. case is analytically tractable, that for the regression setting is more complicated because the normalizing constant $Z(\lambda, \lambda \bm{m})$ in (\ref{eq:glm_jointpost}) does not have a closed form in general. 


We may utilize an integrated Laplace approximation to estimate the normalizing constant. \citet{sabanes_bove_hyper-g_2011} took a similar approach to estimating the marginal likelihood when $\lambda$ is treated as random. Note that $\CI(\bm{\beta} | a, \bm{b} / a)$ is proportional to the likelihood of a GLM with response variable $\bm{b}$ and inverse dispersion parameter $a$ with link function $\theta$. Thus, we may utilize maximum likelihood methods to efficiently obtain, for each proposed value $\tilde{\bm{m}}$ of $\bm{m}$, $\hat{\bm{\beta}}_{\tilde{\bm{m}}}$, the value of $\bm{\beta}$ that maximizes $\CI(\bm{\beta} | 1 + \lambda, (\bm{y} + \lambda \bm{m})/(1 + \lambda)$, and $\mathcal{J}(\hat{\bm{\beta}}_{\tilde{\bm{m}}})$, the observed information matrix evaluated at the maximizer. A Laplace approximation to the normalizing constant $Z(\lambda, \lambda \bm{m})$ of the prior (\ref{eq:glm_conjprior}) is given by
\begin{align*}
    \hat{Z}_{\text{L}}(\lambda, \lambda \bm{m}) \equiv (2\pi)^{1/2} \lvert \lambda \mathcal{J}( \hat{\bm{\beta}}_{\bm{m}}) \rvert^{-1/2} \exp\left\{ \lambda \left[ \bm{m}'\theta( \bm{X} \hat{\bm{\beta}}_{\bm{m}} ) - \bm{J}'b( \theta( \bm{X} \hat{\bm{\beta}}_{\bm{m}} ) ) \right] \right\}.
\end{align*}
Because $\hat{Z}_\text{L}$ is a nonstochastic estimator of the normalizing constant, we may utilize Hamiltonian MCMC for posterior sampling, such as the highly efficient No U-Turn Sampler (NUTS) of \citet{homan_no-u-turn_2014}. 

Under the special case that all covariates are categorical, there are closed-form expressions for the normalizing constants, and sampling from the posterior (\ref{eq:glm_jointpost}) via MCMC is straightforward. Suppose without loss of generality that there is a single categorical covariate with $J$ levels. Let $\bm{y}_j = (y_{1j}, \ldots, y_{n_j,j})$ denote the $n_j$ response variables belonging to the $j^{th}$ category, $j = 1, \ldots, J$. The likelihood may be written as
\begin{align}
L(\bm{\beta} | \bm{y}) \propto \prod_{j=1}^J \exp\left\{ n_j \left[ \beta_j \bar{y}_j - b(\beta_j) \right] \right\},
\label{eq:glm_cat_like}
\end{align}
where $\bar{y}_j = n_j^{-1} \sum_{i=1}^{n_j} y_{ij}$ is the sample mean for the $j^{th}$ category. Note that the likelihood (\ref{eq:glm_cat_like}) is proportional to the product of $J$ exponential family likelihoods (\ref{eq:iid_expfamdens}), where observations are i.i.d. within each category. It follows that the joint conjugate prior of $\bm{\beta}$ is the product of $J$ independent DY priors, i.e.,
\begin{align}
    \pi(\bm{\beta} | \lambda, \bm{m}) \propto \prod_{j=1}^J \frac{1}{Z(\lambda, \lambda m_j)} \exp\left\{ \lambda\left[ m_j \beta_j - b(\beta_j) \right] \right\},
    \label{eq:glm_cat_prior}
\end{align}
where $\bm{m} = (m_1, \ldots, m_J)'$ is a $J$-dimensional vector. Since (\ref{eq:glm_cat_prior}) is a product of independent conjugate priors with a scalar shape parameter, the normalizing constant for each class has a known closed form (e.g., it is a beta function for a logistic regression model).

\section{Utilizing previous studies}
\label{sec:previousstudies}
In this section, we describe how the HPP may be utilized when we possess summary statistics from a previous study obtained, for example, from a publication. We compare and contrast our approach with the power prior of \citet{ibrahim_power_2000}.

Suppose that we possess historical data $D_0 = \{ y_{0i}, i = 1, \ldots, n_0 \}$ with density (or mass function) (\ref{eq:iid_expfamdens}). For notational ease, we assume that dispersion parameters are known and fixed. The power prior of \citet{ibrahim_power_2000} in i.i.d. settings is given by
\begin{align}
    \pi_{\text{PP}}(\theta | a_0, D_0) \propto \exp\left\{ a_0 \left[ \theta \bar{y}_0 - b(\theta) \right] \right\} \pi_0(\theta), \theta \in \Theta,
    \label{eq:iid_powerprior}
\end{align}
where $a_0 \in (0, n]$ is a parameter controlling how much the practitioner wishes to borrow from the historical data and $\pi_0(\theta)$ is an initial prior for $\theta$. If $\pi_0(\theta) \propto 1$, then the power prior (\ref{eq:iid_powerprior}) is precisely the DY prior (\ref{eq:iid_conjprior}) with $\lambda = a_0$ and $m = \bar{y}_0$. Thus, for i.i.d. exponential family models, one may obtain the same posterior density using the summary statistic $S_0 = \bar{y}$ as that using the entire historical data set $D_0$.

Suppose now that we possess historical data with covariates, say, $D_0 = \{ (y_{0i}, \bm{x}_{0i}), i = 1, \ldots, n_0 \}$. The power prior for GLMs is given by
\begin{align}
  \pi_{\text{PP}}(\bm{\beta} | a_0, D_0) \propto \exp\left\{ a_0 \left[ \bm{y}_0' \theta(\bm{X}_0\bm{\beta}) - \bm{J}'b(\theta(\bm{X}_0 \bm{\beta})) \right] \right\} \pi_0(\bm{\beta}), \bm{\beta} \in \mathbb{R}^p,
  \label{eq:glm_powerprior}
\end{align}
where $a_0 \in (0, 1]$ is described above and $\pi_0$ here is an initial prior for $\bm{\beta}$, which we may take as $\pi_0(\bm{\beta}) \propto 1$. The prior (\ref{eq:glm_powerprior}) is similar to CI prior with $\lambda = a_0$, but $\bm{y}_0$ is a $n_0$-dimensional vector while $\bm{m}$ in the CI prior is a $n$-dimensional vector and the covariates from the historical data set, $\bm{X}_0$, are utilized instead of those from the current data, $\bm{X}$. Note that for GLMs, unlike the i.i.d. case, the power prior requires the full historical data set.

We may, however, utilize the MLE from the historical data, $\hat{\bm{\beta}}_0$, to obtain a prior prediction for the CI prior as $\bm{\mu}_0 = g^{-1}(\bm{X}\hat{\bm{\beta}}_0)$. However, $\hat{\bm{\beta}}_0$ is a statistic and each of its components has a variance, and there is no direct mechanism for implementing this uncertainty into the CI prior. 

By contrast, using the HPP, we may elicit uncertainty surrounding the prior prediction $\bm{\mu}_0$ via the precision parameter $\lambda_0$. Using the delta method, we may approximate the variance of $\mu_{0i} = g^{-1}(\bm{x}_i' \hat{\bm{\beta}}_0)$ as
\begin{align}
    \text{Var}(g^{-1}(\bm{x}_i'\hat{\bm{\beta}}_0)) \approx \frac{\bm{x}_i' \text{Cov}(\hat{\bm{\beta}}_0) \bm{x}_i}
    {\left[ \dot{g} \circ g^{-1}(\bm{x}_i'\hat{\bm{\beta}}_{0})\right]^2} \equiv \tau_{0i}, i = 1, \ldots, n.
    \label{eq:glm_deltamethod}
\end{align}
Typically, previous studies report only the estimated standard errors, $\{ \hat{\sigma}_{0j}, j = 1, \ldots, p \}$, of $\hat{\bm{\beta}}_0$ instead of its estimated covariance matrix. In such cases, we may substitute 
$\widehat{\bm{\Sigma}} = \text{diag}\{ \hat{\sigma}_{0j}^2, j = 1, \ldots, p \}$ for $\text{Cov}(\hat{\bm{\beta}}_0)$ in (\ref{eq:glm_deltamethod}), providing a reasonable approximation for $\tau_{0i}$ in (\ref{eq:glm_deltamethod}), say, $\hat{\tau}_{0i}$ as
\begin{align*}
  \hat{\tau}_{0i} = \frac{\sum_{j=1}^p x_{ij}^2 \hat{\sigma}_{0j}^2}
   {\left[ \dot{g} \circ g^{-1}(\bm{x}_i'\hat{\bm{\beta}}_0)\right]^2}, i = 1, \ldots, n.
   %
\end{align*}
Once an estimate $\hat{\tau}_{0i}$ for $\tau_{0i}$ is obtained, we may elicit $\mu_{0i}$ and $\lambda_{0i}$, such that $\text{Var}(m_i) = \hat{\tau}_{0i}$, $i = 1, \ldots, n$. Here, we allow each component of $\bm{m}$ to have its own precision. That is, we augment the hyperprior (\ref{eq:glm_mprior}) to 
\begin{align*}
    \HPP(\bm{m} | \bm{\lambda}_0, \bm{\mu}_0) \propto \prod_{i=1}^n \exp\left\{ \lambda_{0i} \left[ \dot{b}^{-1}(m_{i}) \mu_{0i} - (b \circ \dot{b}^{-1})(m_{i}) \right] \right\} \frac{1}{v(m_i)},
    %
\end{align*}
where $\bm{\lambda}_0 = (\lambda_{01}, \ldots, \lambda_{0n})'$ is a $n$-dimensional vector of precision parameters. We stress that the hyperparameters $\bm{\lambda}_0$ and $\bm{\mu}_0$ in the example described are not elicited based on an individual's opinion. Rather, the hyperparameters are deterministic functions of summary statistics from a previous study.

Both the HPP and the power prior of \citet{ibrahim_power_2000} induce an a priori correlation structure on the regression coefficients. However, with the power prior, the correlation structure depends on the observed historical response variable $\bm{y}_0$ and design matrix $\bm{X}_0$. By contrast, the correlation structure in the HPP is based on the current design matrix $\bm{X}$, the prior prediction $\bm{\mu}_0 = g^{-1}(\bm{X}\hat{\bm{\beta}}_0)$, and the prior precision $\bm{\lambda}_0$. When one is in possession of an entire historical data set, the power prior may be used, which is less computationally demanding and has various desirable properties (see, e.g., \citet{ibrahim_power_2015}). However, the HPP may be used more generally, as it only requires a prior prediction for the mean response and associated uncertainty, which may be obtained either from a full historical data set or using summary statistics from a publication.

When the parameter $a_0$ in the power prior (\ref{eq:glm_powerprior}) is treated as fixed, the prior can be highly influential on the posterior density. For example, in Section \ref{sec:sims}, we generate a historical data set with a positive treatment effect and a current data set with a null treatment effect. For these incompatible data sets, the power prior suggested that treatment was all but certain to be efficacious a posteriori. By contrast, the posterior density of the treatment effect under the HPP suggested that treatment may be unbeneficial, which is the correct result based on how the current data were generated.

\section{Data application}
\label{sec:data}
In this section, we obtain posterior quantities under the HPP using the data of \citet{finney_estimation_1947} (henceforth referred to as the Finney data), which was obtained from the R package \emph{robustbase} \citep{maechler_robustbase_2021}. We compare posterior quantities with the CI prior. In particular, we illustrate how the HPP is less sensitive to the prior prediction than the CI prior.

The Finney data were obtained from a controlled study in which the effects of the rate and volume of air from a single deep breath on whether a transient reflex vaso-constriction on the skin of the digits occurred. The response variable, $y$, is binary, such that $y = 1$ if a vaso-constriction occurred and $y = 0$ if one did not occur. The number of observations recorded is $n = 39$. There are two continuous covariates: $x_1 = \log(\text{volume})$ and $x_2 = \log(\text{rate})$, where ``volume'' is the inhaled volume of the individual and ``rate'' is the rate of inhalation. We assume a logistic regression model for the responses. The likelihood may thus be written as
\begin{align*}
    L(\bm{\beta} | \bm{y}, \bm{X}) = \prod_{i=1}^n \exp\left\{ y_i \bm{x}_i'\bm{\beta} - \log(1 + \exp\{ \bm{x}_i'\bm{\beta} \}) \right\},
\end{align*}
where $\bm{x}_i = (1, x_{1i}, x_{2i})'$, $i = 1, \ldots, n$, $\bm{X} = (\bm{x}_1, \ldots, \bm{x}_n)'$ is the $n \times 3$ design matrix, and $\bm{\beta} = (\beta_0, \beta_1, \beta_2)$, where $\beta_0$ is an intercept term. The CI prior takes the form
$$
    \CI(\bm{\beta} | \lambda, \bm{m}) \propto \prod_{i=1}^n \exp\left\{ \lambda\left[ m_i \bm{x}_i'\bm{\beta} - \log(1 + \exp\{\bm{x}_i'\bm{\beta} \}) \right] \right\},
$$
where $\bm{m} = (m_1, \ldots, m_n)' \in (0, 1)^n$ is the shape parameter and $\lambda \in (0, 1]$ is the precision parameter. The hierarchical model is completed by specifying the hyperprior of the HPP, which is given by
\begin{align}
    \HPP(\bm{m} | \bm{\mu_0}, \lambda_0) \propto \prod_{i=1}^n m_i^{\lambda_0 \mu_{0i}} (1 - m_i)^{\lambda_0 (1 - \mu_{0i})},
    \label{eq:logistic_hyperprior}
\end{align}
where $\bm{\mu} \in (0, 1)^n$ is the prior prediction for the mean of $\bm{y}$ and $\lambda_0 > 0$ is the associated precision parameter. Note that (\ref{eq:logistic_hyperprior}) is proportional to a collection of $n$ independent beta random variables each with mean $\mu_{0i}$ and shared precision parameter $\lambda_0$, where $\mu_{0i}$ denotes the $i^{th}$ component of $\bm{\mu}_0$. 

For the purposes of illustration, we suppose that an expert informs us that lower volume, on average, yields a lower probability for the event. However, the expert is unsure about the effect of the rate of inhalation, and declines to make a prediction for the probability of an event based on that covariate. The expert then gives their prediction based on the level of volume, providing a value of $\mu_{0i} = \mu_0$ when volume is less than 1 and $\mu_{0i} = \mu_1$ when volume is larger than 1, for $i = 1, \ldots, n$. The expert further dictates their uncertainty surrounding these values, which is captured by $\lambda_0 > 0$.

The following hyperparameters for the exercise were used: $(\mu_0, \mu_1) \in \{ (0.3, 0.7), (0.4, 0.6) \}$, $\lambda_0 \in \{ 4, 10 \}$, and $\lambda \in \{ 0.75, 1.00 \}$. We note that the choices for prior mean were arbitrary and they need not be symmetric in general. Models were fit using the \emph{rstan} package in R \citep{stan_development_team_rstan_2020}, which utilizes the No U-Turn Sampler (NUTS) of \citet{homan_no-u-turn_2014}. The Laplace method for estimating the normalizing constant in Section \ref{subsec:computation} was utilized. A total of 25,000 samples were obtained, for which 1,000 was used as a burn-in period. No thinning was performed. The minimum effective sample size (ESS) for all samples of $\bm{\beta}$ across all scenarios was 9,269.

Table~\ref{tab:dataanalysis}, shows the results from the analysis of the Finney data, comparing the posterior under the HPP against the CI prior. The table shows that posterior quantities utilizing the HPP are more similar to those using the CI prior for increasing values of $\lambda_0$. For example, for $(\mu_0, \mu_1) = (0.3, 0.7)$ and when $\lambda = 0.75$, the posterior means for $\beta_1$ are, respectively, $2.72$, $2.63$, and $2.56$ for $\lambda_0$ equal to $4$, $10$, and $\infty$. For example, when $\lambda_0 = 4$, the posterior mean for $\beta_1$ is $2.72$ and $2.56$, respectively, when $\lambda = 0.75$, and it is $2.53$ and $2.35$, respectively, when $\lambda = 1$. A heuristic reason for this is because as $\lambda$ increases, the posterior depends more on the prior. Smaller values of $\lambda_0$ suggest more uncertainty in the prior prediction $\mu_0$. Hence, for fixed values of $\lambda$, lower values of $\lambda_0$ will make the posterior distribution under the HPP more dissimilar to that under the CI prior.

The primary advantage of the HPP over the CI prior is that it enables practitioners to elicit an appropriate amount of uncertainty surrounding the prior prediction. As previously stated, in general, each component of the prior prediction may have its own variance. Moreover, Table~\ref{tab:dataanalysis} clearly shows that the results become similar when the precision parameter $\lambda_0$ increases. However, due to the estimation of the normalizing constant of the conjugate prior in the MCMC algorithm, the HPP is more computationally expensive. Despite this, we were able to obtain a large amount of samples in a matter of a few minutes.

\begin{table}
\caption{\label{tab:dataanalysis}Posterior quantities from Finney data}
\centering
\scalebox{0.75}{
\begin{tabular}{lllcccccc}
\toprule
& & & \multicolumn{6}{c}{$(\mu_0, \mu_1)$} \\ \cmidrule(lr){4-9}
& & & \multicolumn{3}{c}{$(0.3, 0.7)$} & \multicolumn{3}{c}{$(0.4, 0.6)$} \\ \cmidrule(lr){4-6}\cmidrule(lr){7-9}
$\lambda_0$ & $\lambda$ & Parameter & Mean & SD & HPD & Mean & SD & \multicolumn{1}{c}{HPD} \\ 
\midrule
\nopagebreak 4 & \nopagebreak 0.75 & \nopagebreak $\beta_0$  & $-0.858$ & $\phantom{-}0.448$ & $(-1.837, -0.074)$ & $-0.796$ & $\phantom{-}0.437$ & $(-1.741, -0.019)$ \\
 &  & \nopagebreak $\beta_1$  & $\phantom{-}2.721$ & $\phantom{-}0.798$ & $(~~1.308, ~~4.439)$ & $\phantom{-}2.230$ & $\phantom{-}0.714$ & $(~~0.951, ~~3.740)$ \\
 &  & \nopagebreak $\beta_2$  & $\phantom{-}1.403$ & $\phantom{-}0.584$ & $(~~0.431, ~~2.711)$ & $\phantom{-}1.353$ & $\phantom{-}0.575$ & $(~~0.378, ~~2.632)$ \\
 & \rule{0pt}{1.7\normalbaselineskip}1.00 & \nopagebreak $\beta_0$  & $-0.736$ & $\phantom{-}0.390$ & $(-1.560, -0.031)$ & $-0.672$ & $\phantom{-}0.383$ & $(-1.489, ~~0.014)$ \\
 &  & \nopagebreak $\beta_1$  & $\phantom{-}2.529$ & $\phantom{-}0.725$ & $(~~1.211, ~~4.063)$ & $\phantom{-}1.998$ & $\phantom{-}0.650$ & $(~~0.803, ~~3.359)$ \\
 &  & \nopagebreak $\beta_2$  & $\phantom{-}1.188$ & $\phantom{-}0.499$ & $(~~0.330, ~~2.285)$ & $\phantom{-}1.134$ & $\phantom{-}0.500$ & $(~~0.282, ~~2.250)$ \\
\rule{0pt}{1.7\normalbaselineskip}10 & \nopagebreak 0.75 & \nopagebreak $\beta_0$  & $-0.793$ & $\phantom{-}0.412$ & $(-1.668, -0.051)$ & $-0.734$ & $\phantom{-}0.408$ & $(-1.603, ~~0.003)$ \\
 &  & \nopagebreak $\beta_1$  & $\phantom{-}2.631$ & $\phantom{-}0.760$ & $(~~1.262, ~~4.244)$ & $\phantom{-}2.131$ & $\phantom{-}0.692$ & $(~~0.874, ~~3.587)$ \\
 &  & \nopagebreak $\beta_2$  & $\phantom{-}1.291$ & $\phantom{-}0.525$ & $(~~0.398, ~~2.440)$ & $\phantom{-}1.242$ & $\phantom{-}0.528$ & $(~~0.324, ~~2.395)$ \\
 & \rule{0pt}{1.7\normalbaselineskip}1.00 & \nopagebreak $\beta_0$  & $-0.669$ & $\phantom{-}0.368$ & $(-1.455, ~~0.006)$ & $-0.613$ & $\phantom{-}0.360$ & $(-1.370, ~~0.046)$ \\
 &  & \nopagebreak $\beta_1$  & $\phantom{-}2.422$ & $\phantom{-}0.682$ & $(~~1.170, ~~3.826)$ & $\phantom{-}1.900$ & $\phantom{-}0.622$ & $(~~0.761, ~~3.193)$ \\
 &  & \nopagebreak $\beta_2$  & $\phantom{-}1.075$ & $\phantom{-}0.453$ & $(~~0.288, ~~2.068)$ & $\phantom{-}1.032$ & $\phantom{-}0.453$ & $(~~0.240, ~~2.013)$ \\
\rule{0pt}{1.7\normalbaselineskip}$\infty$ & \nopagebreak 0.75 & \nopagebreak $\beta_0$  & $-0.747$ & $\phantom{-}0.396$ & $(-1.592, -0.033)$ & $-0.682$ & $\phantom{-}0.385$ & $(-1.484, ~~0.015)$ \\
 &  & \nopagebreak $\beta_1$  & $\phantom{-}2.563$ & $\phantom{-}0.729$ & $(~~1.228, ~~4.080)$ & $\phantom{-}2.039$ & $\phantom{-}0.658$ & $(~~0.837, ~~3.410)$ \\
 &  & \nopagebreak $\beta_2$  & $\phantom{-}1.212$ & $\phantom{-}0.498$ & $(~~0.352, ~~2.312)$ & $\phantom{-}1.159$ & $\phantom{-}0.492$ & $(~~0.305, ~~2.223)$ \\
 & \rule{0pt}{1.7\normalbaselineskip}1.00 & \nopagebreak $\beta_0$  & $-0.626$ & $\phantom{-}0.341$ & $(-1.331, -0.003)$ & $-0.560$ & $\phantom{-}0.331$ & $(-1.250, ~~0.050)$ \\
 &  & \nopagebreak $\beta_1$  & $\phantom{-}2.351$ & $\phantom{-}0.653$ & $(~~1.154, ~~3.707)$ & $\phantom{-}1.801$ & $\phantom{-}0.579$ & $(~~0.731, ~~2.985)$ \\
 &  & \nopagebreak $\beta_2$  & $\phantom{-}0.998$ & $\phantom{-}0.415$ & $(~~0.256, ~~1.884)$ & $\phantom{-}0.939$ & $\phantom{-}0.415$ & $(~~0.204, ~~1.832)$ \\
\bottomrule 
\end{tabular}
}
\end{table}

The HPP is less sensitive to the prior prediction than the CI prior. For example, suppose now the expert believes that higher levels of volume \emph{reduce} the probability of an event. This expert elicits $(\mu_0, \mu_1) \in \{ (0.6, 0.4), (0.7, 0.3) \}$ with the same levels of precision as before. The $95\%$ highest posterior density (HPD) regions under the prior predictions are presented in Figure 1 of the Supplementary Materials. Under the prior response elicitation least consistent with the data, e.g., for $(\mu_0, \mu_1) = (0.7, 0.3)$, the left endpoint of the HPD interval for $\beta_1$ is only positive when $\lambda = 0.75$ and $\lambda_0 = 4$, i.e., when there is more uncertainty surrounding the prior prediction and when the degree of informativeness of the prior is moderate.

\section{Data application with a previous study}
\label{sec:sims}
In this section, we show results applying the proposed HPP against other priors for two generated historical data sets. The other selected priors are the CI prior, the power prior (PP) of \citet{ibrahim_power_2000}, and a ``Gaussian power prior'' (GPP), which is an approximation to the asymptotic power prior of \citet{ibrahim_power_2015} and is given by
$$
\pi_{\text{GPP}}(\bm{\beta} | \bm{\mu}_{\bm{\beta}}, \bm{\Sigma}_{\bm{\beta}}, \lambda) \propto \left[ (2\pi)^{-1/2} \lvert \bm{\Sigma}_{\bm{\beta}} \rvert^{-1/2} \exp\left\{ -\frac{1}{2} (\bm{\beta} - \bm{\mu}_{\bm{\beta}})' \bm{\Sigma}_{\bm{\beta}}^{-1} (\bm{\beta} - \bm{\mu}_{\bm{\beta}} \right\} \right]^{\lambda},
$$
where $\bm{\mu}_{\bm{\beta}}$ is the prior mean of $\bm{\beta}$ and $\bm{\Sigma}_{\bm{\beta}}$ is the prior covariance matrix of $\bm{\beta}$. We assume for the GPP that only the maximum likelihood estimates and associated standard errors are available from the previous study, and we elicit $\bm{\mu}_{\bm{\beta}} = \hat{\bm{\beta}}$ and $\bm{\Sigma}_{\bm{\beta}} = \text{diag}\{ \hat{\sigma}_j, j = 1, \ldots, p \}$, where $\hat{\bm{\beta}} = (\hat{\beta}_1, \ldots, \hat{\beta}_p)'$ is a vector of MLEs for the regression coefficients and $\hat{\sigma}_j$ is the standard error for the $j^{th}$ coefficient, $j = 1, \ldots, p$. The GPP is precisely equivalent to the asymptotic power prior, except the off-diagonal elements of $\bm{\Sigma}_{\bm{\beta}}$ are set to 0. The reason for this is, in practical situations where a prior is being formulated on the basis of results in published studies, the correlations of the regression coefficients are not typically reported.

The PP will use a full historical data set, while hyperparameter elicitation for the three remaining priors will utilize the maximum likelihood estimates (MLEs) and standard errors. For the HPP, $\bm{\lambda}_0$ will be elicited based on the standard errors as described in Section \ref{sec:previousstudies}. For the GPP, the prior variance of the regression coefficients will be set equal to the squared standard errors.

\subsection{The generated data sets}
The ACTG036 study \citep{merigan_placebo-controlled_1991} was a clinical trial comparing AZT with a placebo in asymptomatic patients with hereditary coagulation disorders (hemophilia). The outcome variable was CD4 count, defined to be the number of CD4 cells (a type of white blood cell that destroys bacteria and fights off infection) per cubic millimeter of blood. Covariates included in the model were treatment ($x_{1i} = 1$ if subject $i$ received AZT, 0 otherwise), race $(x_{2i} = 1$ if subject $i$ was white, 0 otherwise), and age ($x_{3i})$, treated as a continuous variable. We assume a Poisson regression model with canonical link. The likelihood function for the current data is given by
\begin{align*}
    L(\bm{\beta} | \bm{y}, \bm{X}) = \exp\left\{ \sum_{i=1}^n y_i\bm{x}_i'\bm{\beta} - \exp\{ \bm{x}_i'\bm{\beta} \}  \right\},
\end{align*}
where $n = 75$ is the sample size of the current data, $y_i$ is the CD4 count for subject $i$, $\bm{x}_i = (1, x_{1i}, x_{2i}, x_{3i})'$, and $\bm{\beta} = (\beta_0, \beta_1, \beta_2, \beta_3)'$. Henceforth, we denote the current data by $D = \{ (y_i, \bm{x}_i), i = 1, \ldots, n \}$. Historical data of size $n_0 = 50$ was generated from the same model, giving the data set $D_0 = \{ (y_{0i}, \bm{x}_{0i}), i = 1, \ldots, n_0 \}$. The quantities $\beta_0, \beta_2,$ and $\beta_3$ were obtained directly from the MLEs of the ACTG036 data set. For the historical data, the treatment effect was set equal to the MLE of the ACTG036 study, that is, $\beta_{1} \approx 0.048$. Two current data sets were generated: an ``incompatible'' current data set, where $\beta_1 = 0$, and a ``compatible'' data set, where $\beta_1 \approx 0.048$.

\subsection{Results}
\begin{figure}[ht]
    \centering
    \includegraphics[width=0.95\textwidth,height = 0.95\textheight,keepaspectratio]{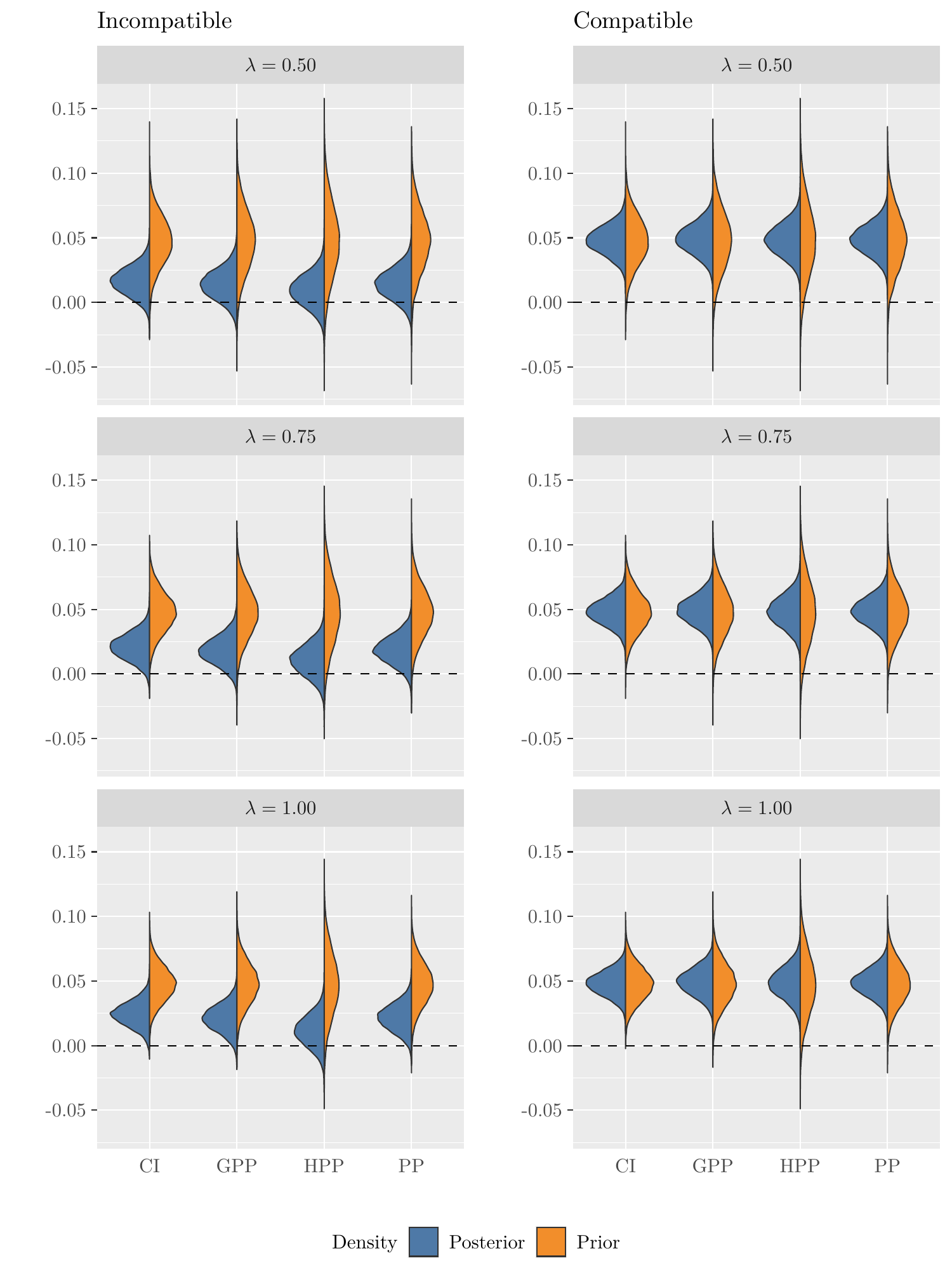}
    \caption{Prior and posterior densities of the treatment effect for the generated data under various levels of borrowing $(\lambda)$. The historical data suggest efficacy ($\beta_1 = 0.048)$. The incompatible current data was generated from $\beta_1 = 0$, while that for the compatible was generated from $\beta_1 = 0.048$.}
    \label{fig:glm_violinplot}
\end{figure}

Prior and posterior density plots for the treatment effect for the compatible and incompatible current data sets are depicted in Figure~\ref{fig:glm_violinplot}. The left panel depicts prior and posterior densities when the historical and current data are incompatible. The right panel shows the prior and posterior densities when the data sets are compatible. Of the four prior densities, the HPP yields the highest prior variance for the treatment effect $\beta_1$ (i.e., it is the least informative prior). When $\lambda \in \{0.75, 1.00\}$, the HPP is the only prior that places mass around the null, while the other three priors suggest that treatment is all but certain to be efficacious a priori. Since the CI prior, which is the most informative prior, is a special case of the HPP, we see that, for a fixed value of $\lambda$, the HPP was the most flexible prior for expressing uncertainty.

When the current data is incompatible with the historical data, i.e., when $\beta_1 = 0$ for the current data and $\beta_1 \approx 0.048$ for the historical data, the posterior distribution of $\beta_1$ utilizing the HPP is generally more robust than its competitors (in the sense of placing the most mass a posteriori on a nonpositive treatment effect). When enabling the prior to have as much influence on the posterior as the data, (i.e., when $\lambda = 1.00$), the posterior densities under the CI prior, GPP, and PP suggest treatment is all but certain to be efficacious a posteriori. In contrast, the posterior density under the HPP suggests that treatment could be unbeneficial, which is the correct result.

When the two data sets are compatible, i.e., when $\beta_1 \approx 0.048$ for both the current and historical data sets, the four posterior densities are quite similar. Across the various levels of $\lambda$, the posterior densities from all four priors suggest that treatment is efficacious a posteriori. The posterior means across the four priors and all levels of $\lambda$ are essentially the same, although the GPP yielded a somewhat higher posterior mean than the other priors. While increasing $\lambda$ reduced the posterior variance markedly for the three competitor priors, the posterior under the HPP was virtually unchanged by the level of $\lambda$.

In general, for this application, the effect of $\lambda$ on the prior and the posterior densities of the treatment effect under the HPP is relatively marginal. When the data sets are incompatible, increasing the value of $\lambda$ from $0.50$ to $0.75$ increases the posterior mean of the treatment effect by $14\%$. By contrast, the same change for the power prior yields roughly a $29\%$ increase, more than twice as much. This suggests that, when each component of the prior prediction for the HPP is given its own level of precision, the posterior density under the HPP is much less sensitive to increasing values of $\lambda$ than the other priors.

The primary advantage of the HPP over the CI prior is that it allows practitioners to incorporate uncertainty in the prior guess in a flexible way. However, the HPP adds computational cost since the normalizing constant of the CI prior must be estimated. The GPP induces an a priori independent prior on the regression coefficients, which is not realistic practically. By contrast, the HPP induces a correlation structure a priori on the regression coefficients. The primary advantage of the PP over the HPP is that the PP is more computationally efficient than the GPP. However, we have seen that the PP can be quite informative when $\lambda$ is fixed, with high posterior probability that $\beta_1$ is positive. By contrast, the posterior under the HPP suggests a nonzero chance that $\beta_1 \le 0$ when the data sets are compatible a posteriori. Moreover, the power prior is somewhat restrictive in that it requires access to a full historical data set.

\subsection{The posterior of m}
In Section~\ref{subsec:mpost}, we argued that the posterior mean of $\bm{m}$ is ``between'' the prior prediction $\bm{\mu}_0$ and the data. We showed that the posterior mean of $\bm{m}$ for the normal linear model was a convex combination of $\hat{\bm{y}}$, the predicted values, and $\bm{\mu}_0$. While a formal proof may not be offered, based on a data set analysis, it appears that, for other GLMs, the posterior mean of $\bm{m}$ is also between $\bm{\mu}_0$ and $\hat{\bm{\mu}} = g^{-1}(\bm{X} \hat{\bm{\beta}})$, the predicted mean based on the MLE.

\begin{figure}
    \centering
    \includegraphics[width = 0.9\textwidth]{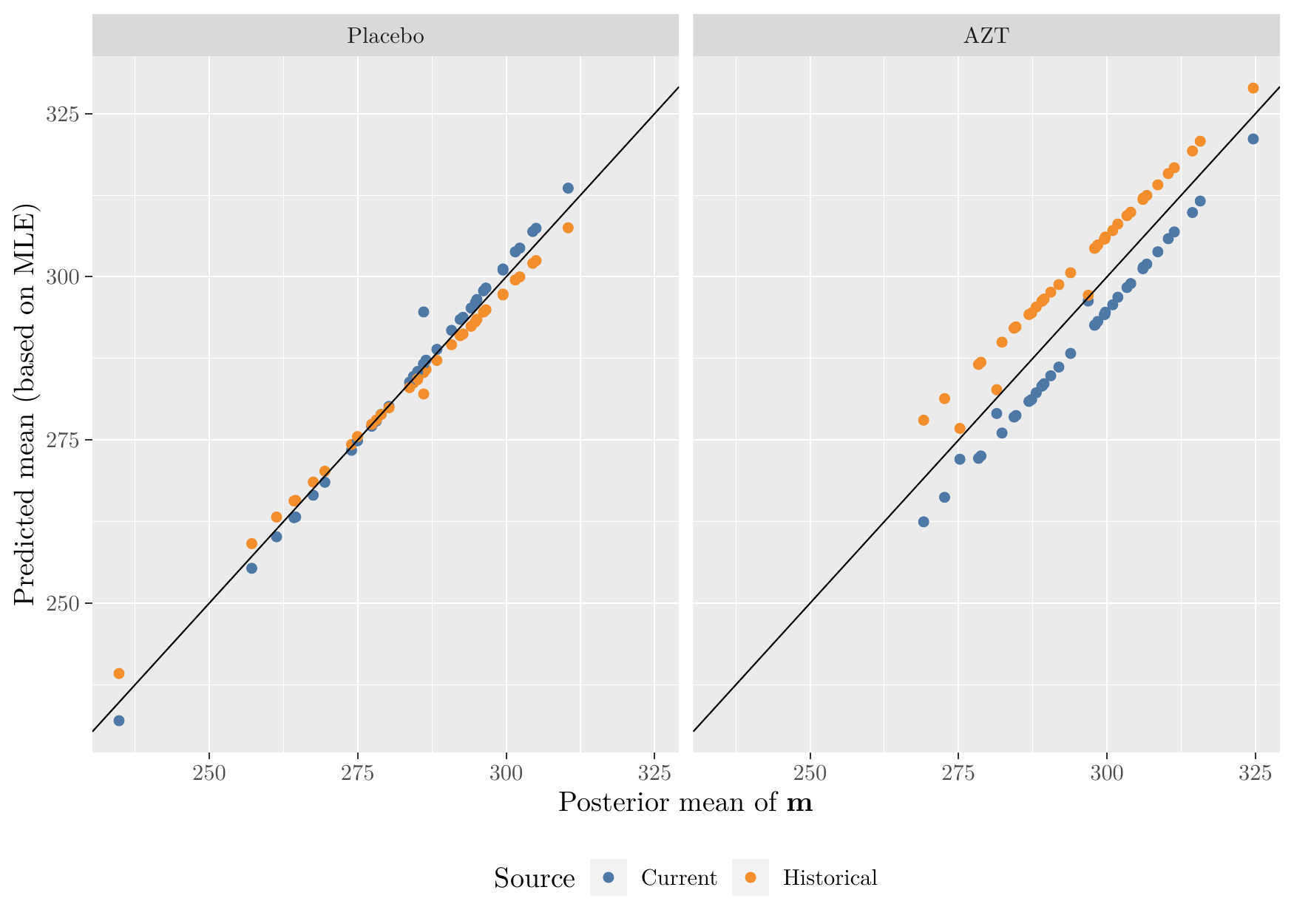}
    \caption{Posterior mean of $\bm{m}$ for the Poisson example in Section \ref{sec:sims} with $\lambda = 1$. The blue points are the predicted mean utilizing the current data set, which suggest $\beta_1 = 0$. The orange points are the predicted mean utilizing the historical data set, which suggest $\beta_1 = 0.048$. The black line is a 45-degree line.}
    \label{fig:mpost_poisson}
\end{figure}

Figure~\ref{fig:mpost_poisson} depicts $\bm{\mu}_0 = g^{-1}(\bm{X}\hat{\bm{\beta}}_0)$, the prediction for $E(\bm{y})$ based on the MLE of the historical data (where $\beta_1 = 0.048)$, and $\hat{\bm{\mu}}$, the predicted mean utilizing the MLE of the incompatible current data (where $\beta_1 = 0$), against the posterior mean of $\bm{m}$. The black line is a 45-degree line, i.e., points close to the black line are close to the posterior mean of $\bm{m}$. 

The left panel of Figure~\ref{fig:mpost_poisson} depicts the predicted means for the historical data set and the posterior of $\bm{m}$ for observations in the control group. For this group, the historical data set is compatible since the covariate corresponding to $\beta_1$ (i.e., the treatment indicator) is zero. Note that the vertical distance between the blue and orange points are relatively small, and both points are close to the 45-degree line. This indicates that the posterior mean of $\bm{m}$ for the control group is very close to both $\hat{\bm{\mu}}$ and $\bm{\mu}_0$, i.e., that $\bm{\mu}_0$ was an accurate guess for $E(\bm{y}_0)$, where $\bm{y}_0$ is the vector of response variables for observations assigned to control.

The right panel of Figure~\ref{fig:mpost_poisson} depicts the posterior mean of $\bm{m}$ against the predicted mean based on the current data (in blue) and historical data (in orange). In this case, the orange points are always above the blue points. This is because, in the current  data, $\beta_1 = 0$, whereas $\beta_1 = 0.048$ in the historical data set (i.e., the predicted mean under the historical data is larger than the current data).


\section{Discussion}
\label{sec:discussion}
We have developed Bayesian hierarchical generalized linear models utilizing conjugate priors, where the mean parameter is treated as random. The HPP is simple and intuitive (e.g., for logistic regression models, the hyperprior is simply a product of independent beta priors, which is, ignoring covariates, a conjugate prior for each component of the response variable). However, we note that the development is flexible--any (proper) hyperprior with support on the mean of the responses may be utilized in practice. For example, a normal distribution truncated to $[0,1]$ may be utilized for logistic regression models.

The incorporation of uncertainty in the prior prediction of the mean of the responses has a natural practical application. As the prior prediction for the mean of the response typically comes in the form of historical data and/or expert opinion, there is uncertainty surrounding the elicited value. Our proposed model enables practitioners to capture this uncertainty. For example, in rare disease settings, the prior prediction may be elicited by a panel of, say, $J$ experts, where $J \ge 1$. The opinions of these experts may be aggregated to a single prior (mathematically or behaviorally), to elicit a predicted response probability $\bm{\mu}_0$ and incorporate uncertainty surrounding this value via $\lambda_0$. We have also explored the relationship of the posterior densities under the HPP and the conjugate prior. 

We have shown several advantages of the HPP over other priors. First, unlike the CI prior, the HPP allows practitioners to fix the degree of influence of the prior while simultaneously allowing for uncertainty surrounding the prior prediction. Second, the HPP may be utilized to set a correlated joint prior on regression coefficients, for example, in settings where summary statistics from a previous study are available, but the full historical data set is not. By contrast, the power prior \citep{ibrahim_power_2000} requires the possession of a full historical data set. Finally, the HPP may be less informative than the CI prior, power prior, and GPP, which may be attractive to clinical regulators. In particular, data applications indicate that posterior quantities using the HPP are robust against prior predictions that are not compatible with the observed data.


An extension of this work could be treating both $\bm{m}$ and $\lambda$ in the CI prior as random. When $\bm{m}$ is fixed and $\lambda$ is random, where $\lambda$ controls for the level of borrowing from the prior. If the prior prediction is incompatible with the data set, the posterior of $\lambda$ would be near zero. However, when $\bm{m}$ and $\lambda$ are both treated as random, it is not clear what the effect would be on the posterior. Another avenue of exploration is an HPP for time-to-event data. While the idea of eliciting a prior prediction for the mean survival time is straightforward, it is not clear how to handle right censored data in such a prior. In longitudinal data, the development of an HPP may be particularly useful since $\bm{m}$ could be elicited as a mean prediction within each cluster.

\bibliographystyle{unsrtnat}
\bibliography{references}  

\end{document}


\if1\blind
{
  \title{\bf Supplementary appendix for: A hierarchical prior for generalized linear models based on predictions for the mean response}
  \author{Ethan M. Alt\thanks{
    This research was supported in part by a grant from the National Institute of Environmental Health Sciences (T32ES007018).}\hspace{.2cm}
    , Matthew A. Psioda, and Joseph G. Ibrahim \\
      Department of Biostatistics, University of North Carolina at Chapel Hill
    }
  \maketitle
} \fi

\if0\blind
{
  \bigskip
  \bigskip
  \bigskip
  \begin{center}
    {\LARGE\bf  Supplementary appendix for: A hierarchical prior for generalized linear models based on predictions for the mean response}
\end{center}
  \medskip
} \fi

\spacingset{1.45} 
\section{The limiting posterior density of $m$}
\label{file2:mpost}
The posterior density of $m$ is
\begin{align*}
    p(m | \bm{y}) &= \int p(\theta, m | \bm{y}) d\theta, \\
                  &\propto \pi(m) \int \frac{1}{Z(\lambda, \lambda m)} \exp\left\{ (n + \lambda) \left[ \frac{n\bar{y} + \lambda m}{n + \lambda} \theta - b(\theta) \right] \right\} d\theta, \\
                  &= \pi(m | \lambda_0, \mu_0) \frac{Z(n + \lambda, n\bar{y} + \lambda m)}{Z(\lambda, \lambda m)},
\end{align*}
where $Z(a, c) = \int \exp\{ c \left[ \theta(c/a) - b(\theta) \right] \} d\theta$.

\subsection{Approximation to a ratio of gamma functions}
For $x$ large,
\begin{align}
    \frac{\Gamma(x + a)}{\Gamma(x)} \sim x^{a} \left[1 + \frac{a(a-1)}{2x} + O(x^{-2}) \right] \approx x^a + \frac{a(a-1)}{2} x^{a-1} + O(x^{a-2})
    \label{eq:gammafunapprox}
\end{align}
This result is presented in, for example, \citet{laforgia_asymptotic_2012}.

\subsection{Poisson model}
Let $y_1, \ldots, y_n$ be i.i.d. observations from $\text{Poisson}(\mu)$. It can be shown that $\HPP(m | \lambda_0, \mu_0) \propto m^{\lambda_0 \mu_0 - 1}e^{-\lambda_0 m}$, so that the hyperprior is a gamma prior with shape parameter $\lambda_0 \mu_0$ and inverse scale parameter $\lambda_0$. Using the marginal posterior (\ref{eq:iid_mpost}) and ignoring constant terms, we may write the posterior for $m$ for the Poisson model as
\begin{align}
p(m | \bm{y}, \lambda, \lambda_0, \mu_0) \propto 
    \frac{\Gamma(\lambda m + n \bar{y})\lambda^{\lambda m}}{\Gamma(\lambda m) (n + \lambda)^{\lambda m} } 
    m^{\lambda_0 \mu_0 - 1} \exp\{ -\lambda_0 m \}.
    \label{eq:poisson_mpost}
\end{align}
An asymptotic approximation for a ratio of gamma functions (see, e.g., \citet{laforgia_asymptotic_2012})  is given by
\begin{align}
    \frac{\Gamma(x + a)}{\Gamma(x + b)} = x^{a - b} \left[1 + \frac{(a - b)(a + b - 1)}{2x} + O(x^{-2}) \right].
    \label{eq:gamma_approximation}
\end{align}
Note that
\begin{align}
    \left( \frac{\lambda}{n + \lambda} \right)^{\lambda m} = \exp\left\{ - \lambda m \log \left( \frac{n + \lambda}{\lambda} \right) \right\}.
    \label{eq:lambda_to_exp}
\end{align}
Utilizing (\ref{eq:lambda_to_exp}) and the approximation (\ref{eq:gamma_approximation}) for the ratio of gamma functions in (\ref{eq:poisson_mpost}), for large values of $\lambda$, we have 
\begin{align*}
    p(m | \bm{y}, \lambda, \lambda_0, \mu_0) &\appropto \left[ (\lambda m)^{n \bar{y}} + \frac{n\bar{y}(n\bar{y} - 1)}{2} (\lambda m)^{n \bar{y} - 1} \right]  m^{\lambda_0 \mu_0 - 1} \exp\left\{ -\left( \lambda_0 + \lambda \log\left( \frac{n + \lambda}{\lambda} \right) \right) m \right\}, \notag
    \\ &\propto \gamma \HPP\left(m | n + \lambda_0, \frac{n\bar{y} + \lambda_0 \mu_0}{n + \lambda_0} \right) + (1 - \gamma) \HPP\left(n + \lambda_0, \frac{n\bar{y} + \lambda_0 \mu_0 - 1}{n + \lambda_0} \right),
\end{align*}
where
$$
  \gamma = \frac{1}{1 + \lambda^{-1} \frac{n\bar{y}(n\bar{y} - 1)}{2}},
$$
where we assume $n\bar{y} > 1$ (i.e., the sum of the Poisson counts are at least 2). Hence, when $\lambda$ is large, we may approximate the posterior of $m$ with a mixture of gamma distributions that have similar shape parameters. Note that as $\lambda \to \infty$, $\gamma \to 1$ and we thus have
$$
  \lim_{\lambda \to \infty} p(m | \bm{y}, \lambda, \lambda_0, \mu_0) \to 
    \HPP\left(m | n + \lambda_0, \frac{n\bar{y} + \lambda_0 \mu_0}{n + \lambda_0} \right),
$$
proving the claim. \\

\subsection{Bernoulli model}
The likelihood for a Bernoulli model is
\begin{align*}
    L(\mu | \bm{y}) = \prod_{i=1}^n \mu^{y_i}(1-\mu)^{1 - y_i} = \mu^{n\bar{y}} (1 - \mu)^{n(1 - \bar{y})}
\end{align*}

The DY prior is given by
\begin{align*}
    \DY(\mu | m, \lambda) = \frac{1}{B(\lambda m, \lambda(1 - m))} \mu^{\lambda m - 1} (1 - \mu)^{\lambda(1 - m) - 1}
\end{align*}

The hyperprior is given by
\begin{align*}
    \HPP(m | \mu_0, \lambda_0) \propto m^{\lambda_0 \mu_0 - 1} (1 - m)^{\lambda_0 (1 - \mu_0) - 1}
\end{align*}
The joint posterior is thus
\begin{align*}
    p(\mu, m | \bm{y}) \propto \frac{1}{B(\lambda m, \lambda(1 - m))} \mu^{n \bar{y} + \lambda m} (1 - \mu)^{n \bar{y} + \lambda(1 - m)} m^{\lambda_0 \mu_0 - 1} (1 - m)^{\lambda_0 (1 - \mu_0) - 1}
\end{align*}

The marginal posterior of $m$ may be written as
\begin{align*}
    p(m |\bm{y}) &\propto \frac{B(\lambda m + n \bar{y}, \lambda(1 - m) + n(1 - \bar{y}))}{B(\lambda m, \lambda(1 - m))} m^{\lambda_0 \mu_0 - 1} (1 - m)^{\lambda_0(1 - \mu_0) - 1} \\
    &\propto  \frac 
    { \Gamma(\lambda_0(1 - m) + n(1 - \bar{y})) \Gamma(\lambda_0 m + n \bar{y}) } 
    { \Gamma(\lambda_0 m) \Gamma(\lambda_0(1 - m)) }  m^{\lambda_0 \mu_0 - 1} (1 - m)^{\lambda_0 (1 - \mu_0) - 1}
    \\ \notag 
    \\ &\approx \left[ (\lambda m)^{n\bar{y}} + \frac{n\bar{y}(n\bar{y}-1)}{2} (\lambda m)^{n\bar{y} - 1} \right]
    \\ & \hspace{.5cm} \times\left[ (\lambda (1-m))^{n(1-\bar{y})} + \frac{n(1-\bar{y})(n(1-\bar{y})-1)}{2} (\lambda (1-m))^{n(1-\bar{y}) - 1} \right]
    \\ & \hspace{.5cm} \times m^{\lambda_0 \mu_0 - 1} (1 - m)^{\lambda_0(1 - \mu_0) - 1 }
    \\ \notag 
    \\ &= \left\{ \left[ \lambda^{n} m^{n\bar{y}} (1-m)^{n(1 - \bar{y})} \right] \right.
    \\ &\hspace{0.5cm} + \left[ \lambda^{n - 1} \frac{n(1 - \bar{y})[n(1 - \bar{y}) - 1]}{2} m^{n\bar{y}} (1 - m)^{n(1 - \bar{y}) - 1} \right]
    \\ &\hspace{0.5cm} + \left[ \lambda^{n-1} \frac{n \bar{y}[n\bar{y}-1]}{2} m^{n\bar{y} - 1} (1 - m)^{n(1 - \bar{y})} \right]
    \\ &\hspace{0.5cm} + \left. \left[ \lambda^{n-2} \frac{n\bar{y}[n\bar{y}-1] n(1 - \bar{y})[n(1-\bar{y}) - 1]}{4} m^{n\bar{y} - 1} (1 - m)^{n(1-\bar{y}) - 1} \right] \right\}
    \\ &\hspace{0.5cm} \times m^{\lambda_0 \mu_0 - 1} m^{\lambda_0 (1 - \mu_0) - 1}
    \\ \notag
    \\ &\propto \pi_1 \text{Beta}(m | \alpha^*, \beta^* ) + \pi_2 \text{Beta}(m | \alpha^* - 1, \beta^* ) + \pi_3 \text{Beta}(m | \alpha^*, \beta^* - 1) 
    \notag \\ & \hspace{0.5cm}+ \pi_4\text{Beta}(m | \alpha^* - 1, \beta^* - 1)
\end{align*}
where $\alpha^* = n\bar{y} + \lambda_0 \mu_0$ and $\beta^* = n(1 - \bar{y}) + \lambda_0(1 - \mu_0)$. Hence, the posterior density of $m$ may be approximated by a four-component mixture of beta densities, where the mixing components depend on the number of successes and failures, and where we used the approximation (\ref{eq:gammafunapprox}).

Note that as $\lambda \to \infty$, $\pi_1 \to 1$. Hence,
\begin{align}
    \lim_{\lambda \to \infty} p(m | \bm{y}) \to \text{Beta}(m | n\bar{y} + \lambda_0 \mu_0, n(1 - \bar{y}) + \lambda_0(1 - \mu_0)).
\end{align}

\subsection{Gamma model}
Assume $\tau = 1$ is known and fixed. Then the likelihood is proportional to
\begin{align*}
    L(\mu | \bm{y}) \propto \mu^{-n} \exp\left\{ -\frac{n \bar{y}}{\mu} \right\}.
\end{align*}
We elicit a conjugate prior with mean $m$ as
\begin{align*}
    \DY(\mu | \lambda, m) &= \frac{(\lambda m)^{\lambda + 1}}{\Gamma(\lambda + 1)} \mu^{-(\lambda + 1) - 1} \exp\left\{ -\frac{\lambda m}{\mu} \right\}.
\end{align*}

The hyperprior for the gamma model is also inverse-gamma
\begin{align*}
    \HPP(m | \lambda_0, \mu_0) &\propto m^{-(\lambda_0 + 1) - 1} \exp\left\{ -\frac{\lambda_0 \mu_0}{m} \right\}.
\end{align*}

The joint posterior is thus
\begin{align*}
    p(\mu, m | \bm{y}) &\propto 
    m^{\lambda - \lambda_0 - 1} \exp\left\{ - \frac{\lambda_0 \mu_0}{m} \right\}
    \mu^{- (n + \lambda + 1) - 1} \exp\left\{ -\frac{n \bar{y} + \lambda m }{\mu}  \right\}
\end{align*}

It follows that
\begin{align*}
    p(m | \bm{y}) &\propto \left( n\bar{y} + \lambda m \right)^{-(n + \lambda + 1)} m^{\lambda - \lambda_0} \exp\left\{ -\frac{\lambda_0 \mu_0}{m} \right\} \\
    &\propto \left( 1 + \frac{\lambda}{n\bar{y}} m \right)^{-(n + \lambda + 1)} m^{-(\lambda_0 - \lambda) - 1} \exp\left\{ -\frac{\lambda_0 \mu_0}{m} \right\} \\
    &= \left[ \frac{\lambda}{n \bar{y}} m \left( 1 + \frac{n\bar{y} / \lambda}{m} \right) \right]^{-(n + \lambda + 1)} m^{-(\lambda_0 - \lambda) - 1} \exp\left\{ -\frac{\lambda_0 \mu_0}{m} \right\} \\
    &\propto \left( 1 + \frac{n\bar{y} / \lambda}{m} \right)^{-(n + \lambda + 1)} m^{-(n + \lambda_0 + 1) - 1} \exp\left\{ -\frac{\lambda_0 \mu_0}{m} \right\}
\end{align*}

For large values of $\lambda$, we have
$$
\left(1 + \frac{n \bar{y} / \lambda}{m} \right)^{-(n + \lambda + 1)} \approx \exp\left\{ - \frac{n\bar{y}}{m} \right\}
$$
so that
\begin{align*}
    p(m | \bm{y}) &\appropto m^{-(n + \lambda_0 + 1)} \exp\left\{ -\frac{n\bar{y} + \lambda_0 \mu_0}{m} \right\}.
    \label{eq:gamma_approx}
\end{align*}
Equation (\ref{eq:gamma_approx}) suggests that, when $\lambda$ is large, the posterior of $m$ may be approximated by an inverse-gamma distribution with shape parameter $n + \lambda_0 + 1$. Hence, when $\lambda$ is large,
$$
  E(m | \bm{y}) \approx \frac{n\bar{y} + \lambda_0 \mu_0}{n + \lambda_0}
$$
\\
\noindent \textbf{Remark:} \\
We may obtain the normalizing constant in terms of studied mathematical functions. Let $Q$ denote the normalizing constant of $p(m | \bm{y})$ when $\lambda_0 > \lambda$. Let $q = m^{-1}$ and let $\tilde{p}$ denote the unnormalized posterior density. Then
\begin{align*}
     \int_\mathbb{R} \tilde{p}(q | \bm{y}) &= \int_{\mathbb{R}} \left(1 + \frac{n\bar{y}}{\lambda} q \right)^{-(n + \lambda + 1)} q^{n + \lambda_0 - 1} \exp\left\{ -(\lambda_0\mu_0) q\right\} dq \\
     &= \frac{\lambda}{n\bar{y}} \int_{\mathbb{R}} \left( 1 + u \right)^{-(n + \lambda + 1)} \left( \frac{\lambda}{n \bar{y}} u \right)^{n + \lambda_0 - 1} \exp\left\{ -\frac{\lambda(\lambda_0 \mu_0)}{n\bar{y}} u\right\} du \\
     &= \left( \frac{\lambda}{n\bar{y}} \right)^{n + \lambda_0} \int_{\mathbb{R}} (1 + u)^{-(n + \lambda + 1)} u^{n + \lambda_0 - 1} \exp\left\{ -\frac{\lambda(\lambda_0 \mu_0)}{n\bar{y}} u\right\} du \\
     &= \left( \frac{\lambda}{n\bar{y}} \right)^{n + \lambda_0} \Gamma(n + \lambda_0) 
       \left[ 
         \frac{1}{\Gamma(n + \lambda_0)} \int_{\mathbb{R}} (1 + u)^{(\lambda_0 - \lambda) - (n + \lambda_0) - 1} u^{n + \lambda_0 - 1} \exp\left\{ -\frac{\lambda(\lambda_0 \mu_0)}{n\bar{y}} u\right\} du 
     \right] \\
     &= \left( \frac{\lambda}{n\bar{y}} \right)^{n + \lambda_0} \Gamma(n + \lambda_0) U(n + \lambda_0, \lambda_0 - \lambda, (\lambda \lambda_0 \mu_0) / (n\bar{y}))
\end{align*}
where
$$
U(a, b, z) = \frac{1}{\Gamma(a)} \int_0^{\infty} e^{-zt} t^{a-1} (1 + t)^{b - a - 1}, a > 0, b > 0, z > 0
$$
is the confluent hypergeometric function of the second kind (also known as Tricomi's confluent hypergeometric function).

\subsection{Normal model}
Suppose $y_1, \ldots, y_n$ are i.i.d. observations from a normal distribution with mean $\mu$ and variance $\sigma^2$. For ease of exposition, we assume $\sigma^2 = 1$ is known. The likelihood of $\mu$ is given by
\begin{align}
    L(\mu | \bm{y}) \propto \exp\left\{ -\frac{n}{2} (\bar{y} - \mu)^2 \right\}.
\end{align}

The conjugate prior for $\mu$ is
\begin{align}
    \DY(\mu | \lambda, m) \propto \exp\left\{ -\frac{\lambda}{2} ( \mu^2 - 2\mu m ) \right\}.
\end{align}
The normalizing constant for the conjugate prior is thus
\begin{align}
    Z(\lambda, \lambda m) &= \int \exp\left\{ -\frac{\lambda}{2} ( \mu^2 - 2\mu m ) \right\} d\mu 
    \notag \\&= (2\pi)^{1/2} \exp\left\{ \frac{\lambda}{2} m^2 \right\} \int (2\pi)^{-1/2} \exp\left\{ -\frac{\lambda}{2} (\mu^2 - 2\mu m + m^2 \right\} d\mu 
    \notag \\&= (2\pi)^{1/2} \exp\left\{ \frac{\lambda}{2} m^2 \right\}
\end{align}


The hyperprior of the HPP is given by
\begin{align*}
    \HPP(m | \lambda_0, \mu_0) \propto \exp\left\{ -\frac{\lambda_0}{2} (m - \mu_0)^2 \right\}
\end{align*}

The joint posterior density is given by
\begin{align*}
    &p(\mu, m | \bm{y}, \lambda, \lambda_0, \mu_0) \notag 
    \\&\propto \exp\left\{ -\frac{1}{2} \left[n \bar{y}^2 - 2\mu n\bar{y} + n\mu^2 + \lambda\mu^2 - 2\mu \lambda m + \lambda m^2 + \lambda_0 m^2 - 2m \lambda_0 \mu_0 + \lambda_0 \mu_0^2 \right] \right\}
    \\&\propto \exp\left\{ -\frac{1}{2} \left[ \mu^2(n + \lambda) - 2\mu (n\bar{y} + \lambda m) + m^2(\lambda + \lambda_0) - 2m \lambda_0 \mu_0 \right] \right\}
    \\&= \exp\left\{ -\frac{n + \lambda}{2} \left[ \mu^2 - 2 \mu \frac{n\bar{y} + \lambda m}{n + \lambda} \right] \right\} 
         \exp\left\{ -\frac{1}{2} \left[ m(\lambda + \lambda_0) - 2m(\lambda_0 \mu_0) \right] \right\}
    \\&= \exp\left\{ -\frac{n + \lambda}{2} \left[ \mu^2 - 2 \mu \frac{n\bar{y} + \lambda m}{n + \lambda} + \left(\frac{n\bar{y} + \lambda m}{n + \lambda}\right)^2 -  \left(\frac{n\bar{y} + \lambda m}{n + \lambda}\right)^2\right] \right\} \notag
    \\ &\hspace{0.5cm} \times \exp\left\{ -\frac{1}{2} \left[ m(\lambda + \lambda_0) - 2m(\lambda_0 \mu_0) \right] \right\}
    \\&\propto \phi\left( \mu \left| \frac{n\bar{y} + \lambda m}{n + \lambda}, \frac{1}{n + \lambda} \right) \right. 
    \times \exp\left\{ -\frac{1}{2} \left[ m^2(\lambda + \lambda_0) - 2m(\lambda_0 \mu_0) -  \frac{(n\bar{y} + \lambda m)^2}{n + \lambda} \right] \right\}
    \\&\propto \phi\left( \mu \left| \frac{n\bar{y} + \lambda m}{n + \lambda}, \frac{1}{n + \lambda} \right) \right. 
    \times \exp\left\{ -\frac{1}{2} \left[ m^2\left( \lambda + \lambda_0 - \frac{\lambda^2}{n + \lambda} \right) - 2m \left(\lambda_0 \mu_0 + \frac{\lambda n \bar{y}}{n + \lambda} \right) \right] \right\}
    \\&= \phi\left( \mu \left| \frac{n\bar{y} + \lambda m}{n + \lambda}, \frac{1}{n + \lambda} \right) \right. 
    \times \exp\left\{ -\frac{1}{2} \left[ \left( \lambda_0 + \frac{n}{1 + n/\lambda} \right) m^2 - 2 m \left(\lambda_0 \mu_0 + \frac{n \bar{y}}{1 + n / \lambda} \right) \right] \right\}
    \\&= \phi\left( \mu \left| \frac{n\bar{y} + \lambda m}{n + \lambda}, \frac{1}{n + \lambda} \right) \right. 
    \times \exp\left\{ -\frac{\left( \lambda_0 + \frac{n}{1 + n/\lambda} \right)}{2} \left[  m^2 - 2 m \frac{\left(\lambda_0 \mu_0 + \frac{n \bar{y}}{1 + n / \lambda} \right)}{\left( \lambda_0 + \frac{n}{1 + n/\lambda} \right)} \right] \right\}
    \\&= \phi\left( \mu \left| \frac{n\bar{y} + \lambda m}{n + \lambda}, \frac{1}{n + \lambda} \right) \right. 
         \times 
      \phi\left( m \left| \frac{\left(\lambda_0 \mu_0 + \frac{n \bar{y}}{1 + n / \lambda} \right)}{\left( \lambda_0 + \frac{n}{1 + n/\lambda} \right)} , \left( \lambda_0 + \frac{n}{1 + n/\lambda} \right)^{-1} \right), \right.
    \\&\equiv p(\mu | m, \bm{y}, \lambda, \lambda_0, \mu_0) p(m | \bm{y}, \lambda, \lambda_0, \mu_0)
\end{align*}
where $\phi(\cdot | a, b)$ is the normal density function with mean $a$ and variance $b$. Clearly, $ \lim_{\lambda \to \infty} p(m | \bm{y}, \lambda, \lambda_0, \mu_0) \to \phi\left(m \left| \frac{\lambda_0 \mu_0 + n\bar{y}}{\lambda_0 + n}, \frac{1}{\lambda_0 + n} \right) \right.$, as was to be shown.

\textbf{Remark}: For the normal model, the mean is always a linear combination of the prior prediction $\mu_0$ and the sample size $n$, i.e.,
$$
E(m | \bm{y}) = \gamma \mu_0 + (1 - \gamma) \bar{y},
$$
where $\gamma = \frac{\lambda_0}{\lambda_0 + \frac{n}{1 + n/\lambda}}$. In addition, the conjugacy of the hyperprior always holds for the normal model (i.e., we do not require $\lambda \to \infty$ for the hyperprior to be conjugate).

\section{Robustness of the HPP}
\begin{figure}[h]
    \centering
    \includegraphics[width=0.9\textwidth]{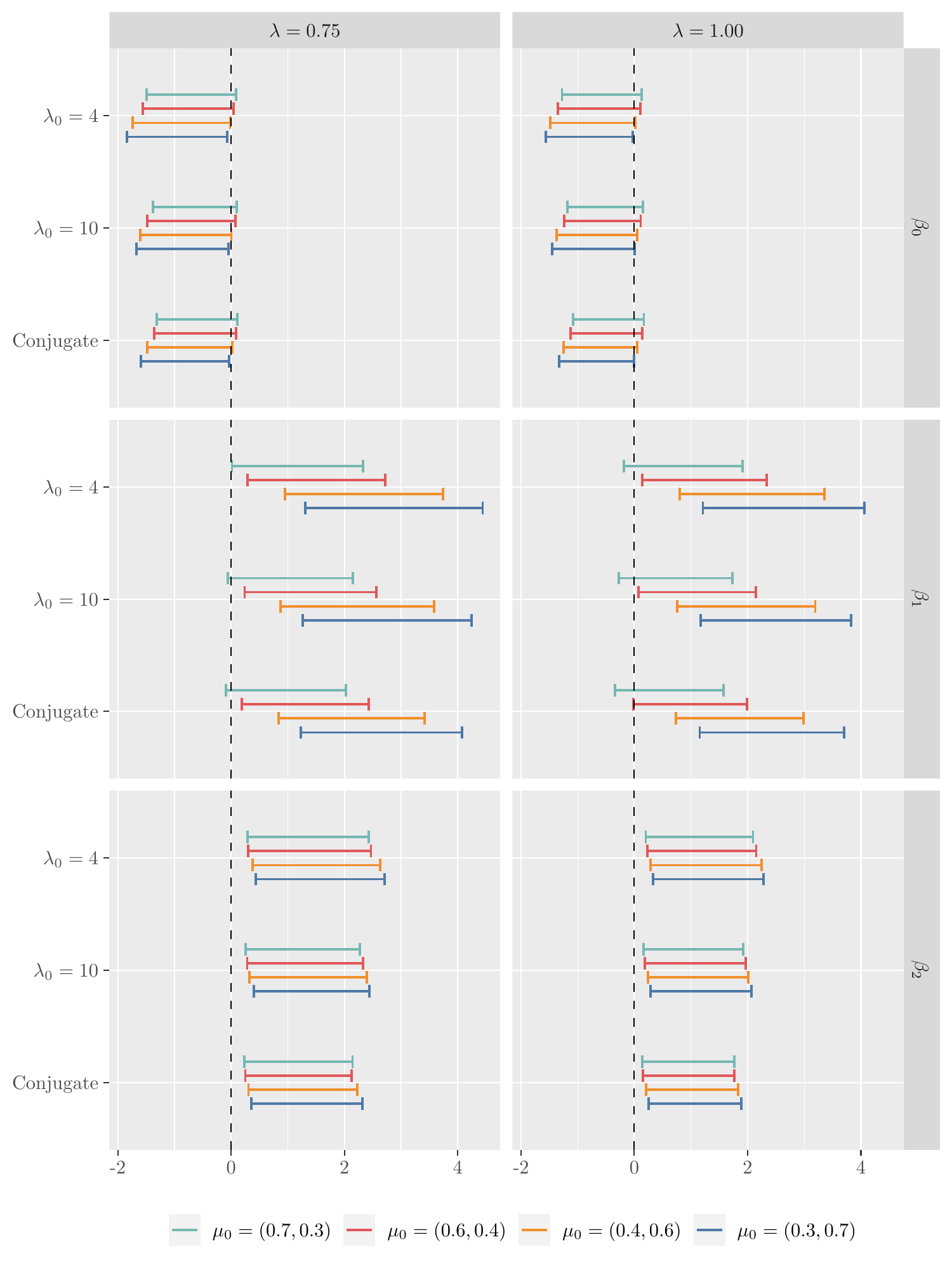}
    \caption{Posterior $95\%$ HPD intervals for regression coefficients under different hyperparameters using the Finney data.}
    \label{fig:finney_forest}
\end{figure}

Suppose now the expert believes that higher levels of volume \emph{reduce} the probability of an event. This expert elicits $(\mu_0, \mu_1) \in \{ (0.6, 0.4), (0.7, 0.3) \}$ with the same levels of precision as before. Figure~\ref{fig:finney_forest} shows $95\%$ highest posterior density (HPD) regions under the prior predictions that were consistent with the data in the previous section and these inconsistent prior predictions. Under the prior response elicitation least consistent with the data, e.g., for $(\mu_0, \mu_1) = (0.7, 0.3)$, the left endpoint of the HPD interval for $\beta_1$ is only positive when $\lambda = 0.75$ and $\lambda_0 = 4$, i.e., when there is more uncertainty surrounding the prior prediction and when the degree of informativeness of the prior is moderate.

 Of course, the data alone suggest that $\beta_1$ should be highly positive. In general, for fixed values of the hyperparameters, the left endpoint of the HPD region of $\beta_1$ is higher than that of the CI prior, suggesting that the proposed HPP is robust against inconsistent predictions of the prior mean response in the sense that it incorporates an appropriate amount of uncertainty, is less informative, and typically results in wider HPD regions than the posterior under the CI prior.

Note that the prior prediction $(\mu_0, \mu_1) = (0.3, 0.7)$ is more informative than when $(\mu_0, \mu_1) = (0.4, 0.6)$ not only because the former has a larger degree of separation between the two constructed groups than the latter, but it also has a lower prior variance for a fixed value of $\lambda_0$. It is interesting to see that the posterior quantities for $\beta_0$ and $\beta_2$ were relatively unaffected by the different hyperparameters. Conversely, the prior prediction $(\mu_0, \mu_1) = (0.3, 0.7)$ substantially lowered both the posterior mean and variance of $\beta_1$, the coefficient associated with volume (the covariate for which the two groups were constructed and the prior prediction was made). By contrast, when the prior prediction is $(\mu_0, \mu_1) = (0.6, 0.4)$, the posterior density under CI prior with $\lambda = 1$ was the only posterior density that suggested the effect of volume could be 0. The widths of the posterior HPD regions under the HPP are all larger than those under the CI prior, reflecting the increased uncertainty in the hyperprior. Interestingly, the difference in widths of the HPD regions for $\beta_1$ when $\lambda = 0.75$ is less for the compatible prediction than for the incompatible prediction.

\section{Posterior distribution of $\bm{m}$ for the normal linear model}
We now derive the marginal posterior distribution of $\bm{m}$ for the normal linear model, We show that the marginal posterior distribution of $\bm{m}$ is multivariate normal. Furthermore, we show that the posterior mean of $\bm{m}$ is a convex combination of the prior prediction $\bm{\mu}_0$ and the predicted values $\hat{\bm{y}} = \bm{X} \hat{\bm{\beta}}$, where $\hat{\bm{\beta}} = (\bm{X}'\bm{X})^{-1} \bm{X}'\bm{y}$ is the MLE of $\bm{\beta}$. Finally, we show that, as $\lambda \to \infty$, the marginal posterior distribution of $\bm{m}$ (and hence of $\bm{\beta}$) is non-degenerate.

Using, from the main text, (\ref{eq:glm_mpost}), (\ref{eq:lm_normconst}), and (\ref{eq:lm_hpp}), we may write the posterior density of $\bm{m}$ as
\begin{align}
    p(\bm{m} | \bm{y}, \lambda, \lambda_0, \bm{\mu}_0) 
    &\propto 
    \frac{\exp\left\{ \frac{1 + \lambda}{2} \left[ \hat{\bm{\beta}}_{\bm{y,m}}' \bm{X}'\bm{X} \hat{\bm{\beta}}_{\bm{y,m}}\right] \right\}}{ \exp\left\{ \frac{\lambda}{2} \left[ \hat{\bm{\beta}}_{\bm{m}}' \bm{X}'\bm{X} \hat{\bm{\beta}}_{\bm{m}}\right] \right\} }
    \exp\left\{ -\frac{\lambda_0}{2} (\bm{m} - \bm{\mu}_0)'(\bm{m} - \bm{\mu}_0) \right\},
    \label{eq:lm_mpost}
\end{align}
where $\hat{\bm{\beta}}_{\bm{y}, \bm{m}} = (\bm{X}'\bm{X})^{-1}\bm{X}'\left( \frac{\bm{y} + \lambda \bm{m}}{1 + \lambda} \right)$ and
$\hat{\bm{\beta}}_{\bm{m}} = (\bm{X}'\bm{X})^{-1}\bm{X}'\bm{m}$. Let $\bm{H} = \bm{X}(\bm{X}'\bm{X})^{-1}\bm{X}'$ denote the orthogonal projection operator onto the column space of $\bm{X}$, often referred to as the ``hat matrix.'' Then 
$\hat{\bm{\beta}}_{\bm{m}}'(\bm{X}'\bm{X})^{-1} \hat{\bm{\beta}}_{\bm{m}} = \bm{m}'\bm{H}\bm{m}$ 
and 
$\hat{\bm{\beta}}_{\bm{y},\bm{m}}'(\bm{X}'\bm{X})^{-1} \hat{\bm{\beta}}_{\bm{y},\bm{m}} 
= 
\frac{1}{(1 + \lambda)^2} (\bm{y} + \lambda \bm{m})'\bm{H} (\bm{y} + \lambda \bm{m})
$. 
It follows that
\begin{align}
    \frac{\exp\left\{ \frac{1 + \lambda}{2} \left[ \hat{\bm{\beta}}_{\bm{y,m}}' \bm{X}'\bm{X} \hat{\bm{\beta}}_{\bm{y,m}}\right] \right\}}{ \exp\left\{ \frac{\lambda}{2} \left[ \hat{\bm{\beta}}_{\bm{m}}' \bm{X}'\bm{X} \hat{\bm{\beta}}_{\bm{m}}\right] \right\} }
    &\propto \frac{\exp\left\{ \frac{1}{2} \left[ \bm{m}' \left( \frac{\lambda^2}{1 + \lambda} \bm{H} \right)\bm{m} + 2 \bm{m}' \frac{\lambda}{1 + \lambda} \hat{\bm{y}} \right] \right\}}{\exp\left\{ \frac{1}{2} \left[ \bm{m}' (\lambda \bm{H}) \bm{m} \right] \right\}},
    \label{eq:lm_normconst_rat}
\end{align}
where $\hat{\bm{y}} = \bm{H}\bm{y} = \bm{X}\hat{\bm{\beta}}$ is the prediction for $E(\bm{y})$ based on MLE $\hat{\bm{\beta}} = (\bm{X}'\bm{X})^{-1}\bm{X}'\bm{y}$. Inserting (\ref{eq:lm_normconst_rat}) into (\ref{eq:lm_mpost}), we get that the posterior of $\bm{m}$ is given by
\begin{align*}
    p(\bm{m} | \bm{y}, \lambda, \lambda_0, \bm{\mu}_0) \propto\exp\left\{ -\frac{1}{2} \left[ \bm{m}' \left( \lambda_0 \bm{I}_n + \frac{\lambda}{1 + \lambda} \bm{H} \right) \bm{m} - 2 \bm{m}' \left( \lambda_0 \bm{\mu}_0 + \frac{\lambda}{1 + \lambda} \hat{\bm{y}} \right) \right] \right\},
    %
\end{align*}
i.e., the posterior distribution of $\bm{m}$ is a $n$-dimensional multivariate normal distribution with positive definite covariance matrix
$\bm{\Sigma}_{\bm{m}} \equiv \left(\lambda_0 \bm{I}_n + \frac{\lambda}{1 + \lambda} \bm{H} \right)^{-1} $ and mean
$\bm{\mu}_{\bm{m}} = \bm{\Sigma}_{\bm{m}} \left( \lambda_0 \bm{\mu}_0 + \frac{\lambda}{1 + \lambda} \hat{\bm{y}} \right)$. The posterior covariance matrix $\bm{\Sigma}_{\bm{m}}$ suggests that smaller values of $\lambda$ yield less dependence between the components of $\bm{m}$ a posteriori. This is intuitive since the components of $\bm{m}$ are independent a priori and, if $\lambda$ is small, the prior has a relatively small impact on the overall posterior.

The posterior mean of $\bm{m}$, $\bm{\mu}_{\bm{m}}$, depends on both the prior prediction $\bm{\mu}_0$ and the fitted values $\hat{\bm{y}}$. In fact, $\bm{\mu}_{\bm{m}}$ is a convex combination of $\bm{\mu}_0$ and $\hat{\bm{y}}$. To see this, note that $\hat{\bm{y}} = \bm{Hy} = \bm{H}(\bm{Hy}) = \bm{H}\hat{\bm{y}}$, where the penultimate equality follows by the idempotency of projection matrices. Thus, we may write
\begin{align*}
\bm{\mu}_{\bm{m}} = \left(\lambda_0 \bm{I}_n + \frac{\lambda}{1 + \lambda} \bm{H} \right)^{-1}\left( \lambda_0 \bm{\mu}_0 + \frac{\lambda}{1 + \lambda} \bm{H}\hat{\bm{y}} \right) = \bm{\Lambda} \bm{\mu}_0 + (\bm{I}_n - \bm{\Lambda}) \hat{\bm{y}},
%
\end{align*}
where 
$\bm{\Lambda} = \left(\lambda_0 \bm{I}_n + \frac{\lambda}{1 + \lambda} \bm{H} \right)^{-1} \lambda_0 \bm{I}_n$, 
so that
$\bm{I}_n - \bm{\Lambda} = \left( \lambda_0 \bm{I}_n + \frac{\lambda}{1 + \lambda} \bm{H} \right)^{-1} \frac{\lambda}{1 + \lambda} \bm{H}.
$
The posterior mean of $\bm{m}$ will be closer to $\hat{\bm{\mu}}$ when $\lambda_0$ decreases or $\lambda$ increases, which reflects the robustness of the HPP. For example, if the CI prior is heavily borrowed from, i.e., if $\lambda \approx 1$, then the posterior mean of $\bm{m}$ will be closer to $\hat{\bm{\mu}}$ than $\bm{\mu}_0$. Conversely, if $\lambda \approx 0$, then the posterior mean of $\bm{m}$ will be closer to $\bm{\mu}_0$, but the posterior of $\bm{\beta}$ will not be affected much by the posterior of $\bm{m}$.

Note that as $\lambda \to \infty$, $\bm{\Sigma}_{\bm{m}}$ is a positive definite matrix and $\bm{\mu}_{\bm{m}}$ is a convex combination of the prior prediction $\bm{\mu}_0$ and predicted values $\hat{\bm{y}}$. The posterior distribution of $\bm{m}$ is hence non-degenerate as $\lambda \to \infty$. By contrast, the posterior distribution under the CI prior becomes a point mass at $\hat{\bm{\beta}}_{\bm{m}}$ as $\lambda \to \infty$, which is a generalization of Theorem 2.1.

\bibliographystyle{plainnat}
\bibliography{references}